\DeclareFontFamily{OT1}{msb}{}{}
\DeclareFontShape{OT1}{msb}{m}{n}
 {  <5> <6> <7> <8> <9> <10> gen * msbm
      <10.95><12><14.4><17.28><20.74><24.88>msbm10}{}
\DeclareMathAlphabet{\bubble}{OT1}{msb}{m}{n}

\def\bN{{\bubble N}}
\def\bZ{{\bubble Z}}

\def\bO{{\bubble O}}
\def\bF{{\bubble F}}

\newfont{\bbd}{msbm10 scaled\magstep1}

\documentclass[12pt,russian]{article}
\usepackage[T2A]{fontenc}
\usepackage[cp866]{inputenc}
\usepackage[russian,english]{babel}
\usepackage{amsfonts}
\parskip=\medskipamount
\begin{document}

\def\l#1#2{\raisebox{.0ex}{$\displaystyle
  \mathop{#1}^{{\scriptstyle #2}\rightarrow}$}}
\def\r#1#2{\raisebox{.0ex}{$\displaystyle
\mathop{#1}^{\leftarrow {\scriptstyle #2}}$}}

\newcommand{\p}[1]{(\ref{#1})}

%%%%%%%%%%%%
\newcommand{\sect}[1]{\setcounter{equation}{0}\section{#1}}
%%%%%%%%%%%%
%\renewcommand{\theequation}{\thesection.\arabic{equation}}

\makeatletter
\def\eqnarray{\stepcounter{equation}\let\@currentlabel=\theequation
\global\@eqnswtrue
\global\@eqcnt\z@\tabskip\@centering\let\\=\@eqncr
$$\halign to \displaywidth\bgroup\@eqnsel\hskip\@centering
  $\displaystyle\tabskip\z@{##}$&\global\@eqcnt\@ne
  \hfil$\displaystyle{\hbox{}##\hbox{}}$\hfil
  &\global\@eqcnt\tw@ $\displaystyle\tabskip\z@
  {##}$\hfil\tabskip\@centering&\llap{##}\tabskip\z@\cr}
%\@addtoreset{equation}{section}
%  \def\theequation{\thesection.\arabic{equation}}
\makeatother

\renewcommand{\thefootnote}{\fnsymbol{footnote}}
%\newpage
%\setcounter{page}{0}
%\pagestyle{empty}
%\begin{flushright}
%{Февраль 2002}\\
%{JINR E2-???-2002}\\
%{nlin.SI/0206044}
%\end{flushright}

%\vfill

%%%%%%%%%%%%%%%%%%%%%%%%%%

\begin{center}
{\sl Theoretical and Mathematical Physics\footnote{Translated from
Teoreticheskaya i Matematicheskaya Fizika
{\bf 132} (2002) 222.}\\
{\bf 132} (2002) 1080, nlin.SI/0206044}
\end{center}

\begin{center}
{\LARGE {\bf N=(1$|$1) supersymmetric }}\\[0.3cm]
{\LARGE {\bf dispersionless}}\\[0.3cm]
{\LARGE {\bf Toda lattice hierarchy}}\\[0.5cm]

{}~

{\large V.G. Kadyshevsky$^{(a)}$ and A.S. Sorin$^{(b)}$}

{}~\\
\quad \\

{{Bogoliubov Laboratory of Theoretical Physics,}}\\
{{Joint Institute for Nuclear Research,}}\\
{\em 141980 Dubna, Moscow Region, Russia}~\quad\\
\em {~$~^{(a)}$ Email: kadyshev@jinr.ru}\\
\em {~$~^{(b)}$ Email: sorin@thsun1.jinr.ru}\\

\end{center}

%\vfill

{}~

\centerline{{\bf Abstract}}
\noindent
Generalizing the graded commutator in superalgebras, we propose a new
bracket operation on the space of graded operators with an involution. We
study properties of this operation and show that the Lax representation of
the two-dimensional $N{=}(1|1)$ supersymmetric Toda lattice hierarchy can
be realized via the generalized bracket operation; this is important in
constructing the semiclassical (continuum) limit of this
hierarchy. We construct the continuum limit of the $N{=}(1|1)$ Toda
lattice hierarchy, the dispersionless $N{=}(1|1)$ Toda hierarchy. In this
limit, we obtain the Lax representation, with the generalized graded
bracket becoming the corresponding Poisson bracket on the graded phase
superspace. We find bosonic symmetries of the dispersionless $N{=}(1|1)$
supersymmetric Toda equation.   
\vskip 2mm

\newpage

%\vfill

%{\it {\Large п}освящается 75--летию академика {\Large а}натолия 
%{\Large  }лексеевича
%{\Large л}огунова}

%{}~

\begin{center}
{\LARGE {\bf N=(1$|$1) суперсимметричная }}\\[0.3cm]
{\LARGE {\bf бездисперсионная}}\\[0.3cm]
{\LARGE {\bf решеточная иерархия Тоды}}\\[0.5cm]

{\large {\LARGE в}.{\LARGE г}. {\LARGE к}адышевский$^{(a)}$ и А.{\LARGE с}.
{\LARGE с}орин$^{(b)}$}{}~\\
\quad \\

{{{\Large л}аборатория {\Large т}еоретической {\Large ф}изики им.
{\Large н}.{\Large н}. {\Large б}оголюбова,}}\\
{{Объединенный {\Large и}нститут {\Large я}дерных {\Large и}сследований,}}\\
{\em {141980 Дубна, {\Large м}осковская область, {\Large р}оссия}}~\quad\\
{\em {~$~^{(a)}$ e-mail: kadyshev@jinr.ru}}\\
{\em {~$~^{(b)}$ e-mail: sorin@thsun1.jinr.ru}}

\end{center}

{}~

\centerline{{\bf Аннотация}}
\noindent
{\Large п}редложена скобочная операция на пространстве градуированных
операторов с инволюцией, обобщающая градуированный коммутатор
супералгебр. {\Large п}оказано, что представление {\Large л}акса для двумерной
$N=(1|1)$ суперсимметричной решеточной иерархии Тоды может быть
реализовано как обобщенная скобочная операция, что важно для
построения квазиклассического (непрерывного) предела этой иерархии.
{\Large п}остроен непрерывный предел $N=(1|1)$ решеточной иерархии
Тоды --- бездисперсионная $N=(1|1)$ иерархия Тоды, и получено
его представление {\Large л}акса, где обобщенная градуированная скобка
переходит в соответствующую скобку {\Large п}уассона на градуированном
фазовом суперпространстве. {\Large н}айдены бозонные симметрии
бездисперсионного $N=(1|1)$ суперсимметричного уравнения Тоды.

{}~

{\it {\Large п}освящается 75--летию академика {\Large а}натолия {\Large а}лексеевича
{\Large л}огунова}

\newpage

%%%%%%%%%%%%%%%%%%%%%%%%%%

\pagestyle{plain}
\renewcommand{\thefootnote}{\arabic{footnote}}
\setcounter{footnote}{0}

\section{{\Large в}ведение}

$\quad ~${\Large в} последние несколько десятилетий квантовая теория поля
({\Large ктп}), приняв на вооружение эффективные математические методы,
превратилась в теорию, удовлетворяющую самым строгим
математическим требованиям \cite{blot}. {\Large п}осле возникновения
суперсимметричных {\Large ктп} особое внимание исследователей привлекли
многочисленные проблемы, которые, c одной стороны, представляют
интерес для математической физики, a, с другой стороны, обещают
важные физические приложения.

{\Large в} настоящей работе будет рассмотрено интегрируемое $N=(1|1)$
суперсимметричное обобщение решеточной двумерной бозонной иерархии
Тоды (2DTL иерархии) \cite{ut}, предложенное в \cite{i,t}. {\Large о}но
представляет собой бесконечную систему эволюционных (по двум
бозонным и двум фермионным бесконечным "башням" времен) уравнений
(потоков) для бесконечного набора решеточных бозонных и фермионных
полей и содержит как подсистему $N=(1|1)$ суперсимметричное
интегрируемое обобщение 2DTL уравнения, обозначаемое в дальнейшем
посредством $N=(1|1)$ 2DTL.

{\Large п}озднее в работах \cite{ls1,ols1,vgks} были построены две новые
бесконечные серии фермионных потоков $N=(1|1)$ 2DTL иерархии и,
как следствие, было установлено, что эта иерархия, в
действительности, обладает более высокой симметрией, а именно
$N=(2|2)$ cуперсимметрией. {\Large э}ти потоки, совместно с ранее
известными бозонными потоками $N=(1|1)$ 2DTL иерархии, являются
симметриями $N=(1|1)$ 2DTL уравнения \cite{ls1,ols1,vgks}.
{\Large н}епрерывный предел по шагу решетки последнего уравнения \cite{ss}
представляет собой трехмерное нелинейное уравнение, называемое
непрерывным или бездисперсионным $N=(1|1)$ 2DTL уравнением; в
\cite{ss} также рассматривалось решение соответствующей задачи
Коши.

{\Large х}отя $N=(1|1)$ 2DTL иерархия и бездисперсионное $N=(1|1)$ 2DTL
уравнение известны достаточно давно, {\it проблема построения
непрерывного (квазиклассического) предела по шагу решетки,
играющего здесь роль постоянной {\Large п}ланка, для всех потоков $N=(1|1)$
2DTL иерархии является еще не решенной и достаточно сложной
задачей}\footnote{{\Large в} этой связи заметим, что аналогичные задачи для
дифференциально--разностных уравнений рассматривались в цикле
работ \cite{vgk1,vgk2,vgk3,vgk4,vgk5,vgk7,vgk8,vgk9}.}. Кроме
чисто академического значения этой проблемы, интерес к ее решению
связан с рядом важных физических и математических приложений. {\Large р}ечь
идет, в частности, о квазиклассическом пределе бозонного прообраза
$N=(1|1)$ 2DTL иерархии --- бездисперсионной 2DTL иерархии
\cite{tt1}, представляющей собой объединение потоков 2DTL
иерархии, возникающих в лидирующем приближении квазиклассического
разложения, построенном в работе \cite{tt2} (см. также обзор
\cite{tt3}). {\Large в} качестве иллюстрации, можно привести перечень
возможных приложений бездисперсионной 2DTL иерархии:\\
$\quad ~~~~~~~$1. построение ряда самодуальных вакуумных метрик и метрик \\
$\quad ~~~~~~~~~~$ {\Large э}йнштейна--{\Large в}ейля; \\
$\quad ~~~~~~~$2. теории твисторов;\\
$\quad ~~~~~~~$3. двумерной конформной и топологической теории
поля;\\
$\quad ~~~~~~~$4. двумерной теории струн \\
(см., также \cite{bf,b,p,w,k,dss,hp,tt4,bx,ct,kkk,ak} и
приведенные там ссылки).

{\Large и}мея в виду глубокую связь между 2DTL и $N=(1|1)$ 2DTL иерархиями,
представляется естественным полагать, что и бездисперсионная
$N=(1|1)$ суперсимметричная 2DTL иерархия найдет аналогичные
приложения в суперсимметричных обобщениях перечисленных выше
теорий. {\it {\Large п}оследнее обстоятельство явилось стимулом для
построения бездисперсионной $N=(1|1)$ 2DTL иерархии в данной
работе.}

Далее мы хотели бы особо подчеркнуть, что существующий общий
алгоритм построения квазиклассических асимптотик, использованный в
работе \cite{tt2} для 2DTL иерархии, встречает ряд как формальных,
так и неформальных препятствий при прямолинейном распространении
на случай $N=(1|1)$ 2DTL иерархии. Так, общеизвестно, что в
квазиклассическом пределе все операторы переходят в их символы,
задаваемые на соответствующем фазовом пространстве, а
(анти)коммутаторы в стандартных представлениях {\Large л}акса заменяются
соответствующими скобками {\Large п}уассона. {\Large ч}то же касается предложенного
в работе \cite{i} представления {\Large л}акса для $N=(1|1)$ 2DTL иерархии,
то оно не имеет (анти)коммутаторного вида. {\Large п}оэтому буквальное
применение вышеприведенного качественного рецепта перехода к
квазиклассике кажется затруднительным.

Другой принципиально важной, качественно новой чертой $N=(1|1)$
2DTL иерархии по сравнению с ее бозонным прообразом, усложняющей
рассмотрение, является то, что операторы, входящие в ее
представление {\Large л}акса, задаются на пространстве с двумя
$Z_2$--градуировками, обладая, однако, только одной диагональной
$Z_2$--градуировкой. {\Large п}о этой причине в общем случае даже в
квазиклассическом пределе нельзя ожидать коммутативности символов
этих операторов.

К сказанному можно также добавить и возможные усложнения,
связанные с тем, что фермионные и бозонные поля $N=(1|1)$ 2DTL
иерархии могут иметь не совпадающие квазиклассические асимптотики,
которые необходимо задать самосогласованно. Аналогичная ситуация
имеет место для фермионных и бозонных времен этой иерархии, однако
последние можно легко согласовать, дополнительно привлекая
соображения размерности.

{\Large с}овсем недавно наметился определенный прогресс на пути построения
квазиклассического предела $N=(1|1)$ 2DTL иерархии. Так, в работе
\cite{vgks} было обнаружено, что представление {\Large л}акса для этой
иерархии может быть приведено к коммутаторной форме (уравнение
(39) в \cite{vgks}) посредством введения ряда новых
вспомогательных фермионных постоянных (детали см. \cite{vgks}).
{\Large х}отя получаемое таким способом коммутаторное представление, на
первый взгляд, кажется искусственным, в действительности, в нем
закодирована важная новая обобщенная градуированная скобочная
операция, которая может быть определена в достаточно общих
терминах, чтобы иметь широкий спектр приложений. {\it Мы вводим ее
в рассмотрение в настоящей работе и используем для решения
изучаемой здесь проблемы.}

Оказывается, что именно в терминах данной скобочной операции могут
быть выражены все основные соотношения, определяющие $N=(1|1)$
2DTL иерархию, и для этого не требуется введение никаких
вспомогательных объектов, типа упомянутых выше фермионных
постоянных. {\Large б}олее того, в квазиклассическом пределе именно эта
скобочная операция заменяется соответствующей суперскобкой
{\Large п}уассона на фазовом суперпространстве. Тем самым задается
представление {\Large л}акса для бездисперсионной $N=(1|1)$ 2DTL иерархии.

{\Large с}труктура данной работы такова. {\Large в} параграфе 2 мы вводим обобщенную
градуированную скобочную операцию на пространстве градуированных
операторов с инволюцией, обобщающую градуированный коммутатор для
супералгебр, описываем ее свойства и приводим соответствующие
обобщенные тождества {\Large я}коби. {\Large з}атем мы получаем
представление {\Large л}акса
$N=(1|1)$ 2DTL иерархии, а также все основные определяющие его
соотношения в терминах обобщенной градуированной скобки. Мы
приводим далее явные выражения для потоков $N=(1|1)$ 2DTL иерархии
и бозонных симметрий $N=(1|1)$ 2DTL уравнения, которые
впоследствии используются в параграфе 3 для получения их
бездисперсионных аналогов. {\Large в} параграфе 3 также определяется
квазиклассический предел $N=(1|1)$ 2DTL иерархии и постулируется
соответствующее асимптотическое поведение фермионных и бозонных
полей, параметризующих операторы {\Large л}акса.

{\Large з}атем с помощью этих данных мы вычисляем асимптотическое поведение
всех композитных операторов, входящих в представление {\Large л}акса и
соответствующие ему полевые эволюционные уравнения. Далее мы
получаем непротиворечивые, регулярные лидирующие члены
квазиклассического разложения этих эволюционных уравнений,
являющихся, по определению, потоками бездисперсионной $N=(1|1)$
2DTL иерархии. Тем самым апостериори демонстрируется
самосогласованность наших постулатов, лежащих в основе всех
проведенных вычислений.

{\Large с}ледующий шаг --- моделирование суперскобки {\Large п}уассона на фазовом
суперпространстве, представляющем собой фазовое пространство
бездисперсионной 2DTL иерархии, расширенное одной грассмановой
координатой. {\Large и}, наконец, заменяя операторы {\Large л}акса на их символы, а
обобщенную градуированную скобку на указанную суперскобку {\Large п}уассона
в представлении {\Large л}акса для $N=(1|1)$ 2DTL иерархии и во всех
определяющих его соотношениях, мы убеждаемся прямым вычислением,
что получаемое таким способом операторное представление правильно
воспроизводит построенные нами ранее потоки бездисперсионной
$N=(1|1)$ 2DTL иерархии, т.е. является искомым представлением
{\Large л}акса для последней иерархии. {\Large в} параграфе 4 мы кратко резюмируем
основные результаты, полученные в работе.

\section{N=(1$|$1) 2DTL иерархия}

$\quad$ {\Large в} этом параграфе вводится новая градуированная скобочная
операция и на этой основе предлагается новая форма для
представления {\Large л}акса $N=(1|1)$ 2DTL иерархии.

\subsection{{\Large о}бобщенные градуированные скобки}

$\quad$ {\Large р}ассмотрим пространство операторов ${\bO}_k$ с
градуировкой $d_{{\bO}_k}$ ($d_{{\bO}_k}\in {\bZ}$),
\begin{eqnarray}
d_{{\bO}_1{\bO}_2}= d_{{\bO}_1}+d_{{\bO}_2}, \label{grad}
\end{eqnarray}
и инволюцией $*$,
\begin{eqnarray}
{\bO}^{*(2)}_k = {\bO}_k, \label{inv}
\end{eqnarray}
где здесь и в дальнейшем ${{\bO}_k}^{*(m)}$ обозначает
$m$--кратное действие инволюции $*$ на оператор ${\bO}_k$. {\Large н}а этом
пространстве можно определить
%следующую
обобщенную градуированную скобочную операцию $[..., ...
\}$:
\begin{eqnarray}
\Bigl[ {\bO}_1, {\bO}_2 \Bigr\}:={\bO}_1{\bO}_2
 - (-1)^{d_{{\bO}_1}d_{{\bO}_2}}~{{\bO}_2}^{*(d_{{\bO}_1})}~
{{\bO}_1}^{*(d_{{\bO}_2})},
  \label{brack}
\end{eqnarray}
со следующими легко проверяемыми свойствами:

{\it {\Large с}имметрия}
\begin{eqnarray}
\Bigl[ {\bO}_1, {\bO}_2 \Bigr\}=(-1)^{d_{{\bO}_1}d_{{\bO}_2}+1}
\Bigl[{{\bO}_2}^{*(d_{{\bO}_1})},{{\bO}_1}^{*(d_{{\bO}_2})}
\Bigr\}, \label{prop1}
\end{eqnarray}

{\it Дифференцирование}
\begin{eqnarray}
\Bigl[ {\bO}_1, {\bO}_2{\bO}_3\Bigr\}=
\Bigl[{\bO}_1,{\bO}_2\Bigr\}{\bO}_3 +
(-1)^{d_{{\bO}_1}d_{{\bO}_2}}{{\bO}_2^{*(d_{{\bO}_1})}}
\Bigl[{{\bO}_1}^{*(d_{{\bO}_2})},{\bO}_3 \Bigr\}, \label{prop2}
\end{eqnarray}

{\it Тождества {\Large я}коби}
\begin{eqnarray}
&&(-1)^{d_{{\bO}_1}d_{{\bO}_3}}\Bigl[\Bigl[
{{\bO}_1}^{*(d_{{\bO}_3)}}, {\bO}_2
\Bigr\},{{\bO}_3}^{*(d_{{\bO}_1)}}\Bigr\} \nonumber\\
&+&(-1)^{d_{{\bO}_2}d_{{\bO}_1}} \Bigl[\Bigl[
{{\bO}_2}^{*(d_{{\bO}_1)}},
{\bO}_3\Bigr\},{{\bO}_1}^{*(d_{{\bO}_2)}}\Bigr\} \nonumber\\&+&
(-1)^{d_{{\bO}_3}d_{{\bO}_2}} \Bigl[\Bigl[
{{\bO}_3}^{*(d_{{\bO}_2)}},
{\bO}_1\Bigr\},{{\bO}_2}^{*(d_{{\bO}_3)}}\Bigr\}=0. \label{jacobi}
\end{eqnarray}
{\Large с}оотношения (\ref{prop1}--\ref{jacobi}) обобщают соответствующие
свойства градуированного коммутатора супералгебр {\Large л}и. Мы хотели бы
подчеркнуть, что в частном cлучае, когда инволюция $*$ \p{inv}
действует как тождественное преобразование, скобка \p{brack}
воспроизводит градуированный коммутатор супералгебр {\Large л}и. {\Large в} случае
же нетривиального действия инволюции эта скобка является
нетривиальным обобщением последнего.

\subsection{{\Large п}редставление {\Large л}акса и потоки}

$\quad$ Мы начнем этот раздел с детализации пространства
операторов, их градуировки и инволюции, имеющих отношение к
рассматриваемой здесь $N=(1|1)$ 2DTL иерархии.

Эти операторы могут быть представлены в следующем общем виде:
\begin{eqnarray}
{\bO}_m=\sum^{\infty}_{k=-\infty}
f^{(m)}_{k,j}e^{(k-m){\partial}}, \quad m\in {\bZ}, \label{optoda}
\end{eqnarray}
где параметризующие их функции $f^{(m)}_{2k,j}$
($f^{(m)}_{2k+1,j}$) являются $Z_2$--градуи-\\рованными
%$(d^{'}_{f^{(m)}_{2k,j}}=0,~d^{'}_{f^{(m)}_{2k+1,j}}=1)$
\begin{eqnarray}
d^{~'}_{f^{(m)}_{k,j}}=|k| \quad \mbox{по ~~ модулю} ~~ 2 \label{grad1}
\end{eqnarray}
бозонными (фермионными) решеточными полями ($j \in {\bZ}$), а
оператор $e^{l{\partial}}$ ($l \in {\bZ}$) действует на эти поля
как дискретный решеточный сдвиг
\begin{eqnarray}
e^{l{\partial}} f^{(m)}_{k,j} \equiv f^{(m)}_{k,j+l}
e^{l{\partial}} \label{rule1}
\end{eqnarray}
и обладает другой $Z_2$--градуировкой
\begin{eqnarray}
d_{e^{l{\partial}}}=|l| \quad \mbox{по ~~ модулю}  ~~ 2. \label{grad2}
\end{eqnarray}
Операторы \p{optoda} допускают задание только одной диагональной
$Z_2$--градуировки
\begin{eqnarray}
d_{{\bO}_m}=d^{~'}_{f^{(m)}_{k,j}}+d_{e^{(k-m){\partial}}} =|m|
\quad \mbox{по ~~ модулю}  ~~ 2
 \label{gradop}
\end{eqnarray}
и инволюции
\begin{eqnarray}
{\bO}^{*}_m=\sum^{\infty}_{k=-\infty} (-1)^k
f^{(m)}_{k,j}e^{(k-m){\partial}}. \label{invtoda}
\end{eqnarray}
{\Large в} дальнейшем нам также понадобятся проекции ${({\bO}_m)}_{\pm}$
операторов ${\bO}_m$ \p{optoda}, определяемые как
\begin{eqnarray}
({\bO}_m)_{+}=\sum^{\infty}_{k=m}
f^{(m)}_{k,j}e^{(k-m){\partial}}, \quad
{({\bO}_m)}_{-}=\sum^{m-1}_{k=-\infty}
f^{(m)}_{k,j}e^{(k-m){\partial}}.
 \label{optoda1}
\end{eqnarray}

Операторы {\Large л}акса $L^{\pm}$ $N=(1|1)$ 2DTL иерархии принадлежат
пространству операторов \p{optoda} \cite{i,vgks}
\begin{eqnarray}
L^{+}=\sum^{\infty}_{k=0} u_{k,j}e^{(1-k){\partial}},\quad
u_{0,j}= 1, \quad L^{-}=\sum^{\infty}_{k=0}
v_{k,j}e^{(k-1){\partial}}, \quad v_{0,j}\neq 0 \label{laxs1}
\end{eqnarray}
и имеют градуировку $d_{L^{\pm}}=1$.

Теперь мы располагаем всеми необходимыми данными для того, чтобы
выразить представление {\Large л}акса $N=(1|1)$ 2DTL иерархии в терминах
скобочной операции \p{brack} и тем самым придать ему очень простой
вид:
\begin{eqnarray}
D^{\pm}_n L^{{\alpha}}&=& \mp {\alpha} (-1)^n
\Bigr[{(((L^{\pm})^{n}_{*})}_{-{\alpha} })^{*},
L^{{\alpha}}\Bigl\}, \quad {\alpha} =+,-, \quad n \in {\bN},
\label{laxreprs1}
\end{eqnarray}
где $D^{\pm}_{2n}$ ($D^{\pm}_{2n+1}$) -- бозонные (фермионные)
эволюционные производные.

Для входящих в это представление композитных операторов
$(L^{\pm})^{n}_{*}$ также могут быть получены весьма простые
выражения в терминах операторов {\Large л}акса и скобочной операции
\p{brack}
\begin{eqnarray}
(L^{{\alpha}})^{2n}_{*}:= \Bigl(~\frac{1}{2}~\Bigr[
(L^{{\alpha}})^{*}, (L^{{\alpha}})\Bigl\}~\Bigr)^n, \quad
(L^{{\alpha}})^{2m+1}_{*}:=L^{{\alpha}}~(L^{{\alpha}})^{2n}_{*}.
\label{rule2}
\end{eqnarray}
Как и операторы {\Large л}акса $L^{\pm}$, операторы $(L^{\pm})^{n}_{*}$
принадлежат пространству операторов \p{optoda} и могут быть
представлены в следующем виде:
\begin{eqnarray}
(L^{+})^{m}_{*}:=\sum^{\infty}_{k=0}
u^{(m)}_{k,j}e^{(m-k){\partial}}, \quad u^{(m)}_{0,j}=1, \quad
(L^{-})^{m}_{*}:=\sum^{\infty}_{k=0}
v^{(m)}_{k,j}e^{(k-m){\partial}}, \label{laxs2}
\end{eqnarray}
где $u^{(m)}_{k,j}$ и $v^{(m)}_{k,j}$ ($u^{(1)}_{k,j}\equiv
u_{k,j},~v^{(1)}_{k,j}\equiv v_{k,j}$) --- функционалы исходных
полей $\{u_{k,j},~v_{k,j}\}$. {\Large з}десь важно отметить, что в
представлении {\Large л}акса \p{laxreprs1} $Z_2$--градуировка операторов
$(L^{\pm})^{n}_{*}$ --- $d_{(L^{\pm})^{2n}_{*}}=0$ и
$d_{(L^{\pm})^{2n+1}_{*}}=1$ --- согласована с другой
$Z_2$--градуировкой $d^{~'}_{D^{\pm}_{2n}}=0$ и
$d^{~'}_{D^{\pm}_{2n+1}}=1$, соответствующей статистике
эволюционных производных $D^{\pm}_{n}$.

{\Large и}спользуя скобочные свойства (\ref{prop1}--\ref{jacobi}) и
соотношения (\ref{rule2}) как определения для $(L^{\pm})^{n}_{*}$,
нетрудно получить полезные тождества
\begin{eqnarray}
&&\Bigr[ (L^{{\alpha}})^{2n}_{*},
(L^{{\alpha}})^{2m}_{*}\Bigl\}=0, \nonumber\\
&&\Bigr[ ((L^{{\alpha}})^{2n}_{*})^{*},
(L^{{\alpha}})^{2m+1}_{*}\Bigl\}=0, \quad \Bigr[
(L^{{\alpha}})^{2n+1}_{*},
(L^{{\alpha}})^{2m}_{*}\Bigl\}=0, \nonumber\\
&&\Bigr[ ((L^{{\alpha}})^{2n+1}_{*})^{*},
(L^{{\alpha}})^{2m+1}_{*}\Bigl\}=2(L^{{\alpha}})^{2(n+m+1)}_{*}.
\label{ident}
\end{eqnarray}
Теперь, применяя (\ref{prop1}--\ref{jacobi}) и
(\ref{laxreprs1}--\ref{rule2}), можно вывести уравнения движения
для композитных операторов $(L^{\pm})^{n}_{*}$
\begin{eqnarray}
D^{\pm}_n (L^{{\alpha}})^{m}_{*}&=& \mp {\alpha} (-1)^{nm} \Bigr[
{(((L^{\pm})^{n}_{*})}_{-{\alpha} })^{*(m)},
(L^{{\alpha}})^{m}_{*}\Bigl\}, \label{laxreprs2}
\end{eqnarray}
а также эволюционные уравнения для функционалов
$\{u^{(m)}_{k,j},~v^{(m)}_{k,j}\}$ \p{laxs2}, вытекающие из
найденных уравнений \p{laxreprs2}:
\begin{eqnarray}
D^{+}_n u^{(2m)}_{k,j}&=&\sum^{n}_{p=0}
(u^{(n)}_{p,j}u^{(2m)}_{k-p+n,j-p+n} \nonumber\\
&-&(-1)^{(p+n)(k-p+n)}u^{(n)}_{p,j-k+p-n+2m}u^{(2m)}_{k-p+n,j}),
\nonumber\\ D^{+}_{2n} u^{(2m+1)}_{k,j}&=&\sum^{2n}_{p=0}
((-1)^{p}u^{(2n)}_{p,j}u^{(2m+1)}_{k-p+2n,j-p+2n} \nonumber\\
&-&(-1)^{p(k-p)}u^{(2n)}_{p,j-k+p-2n+2m+1}u^{(2m+1)}_{k-p+2n,j}),
\nonumber\\ D^{+}_{2n+1} u^{(2m+1)}_{k,j}&=&\sum^{k}_{p=1}
((-1)^{p+1}u^{(2n+1)}_{p+2n+1,j}u^{(2m+1)}_{k-p,j-p} \nonumber\\
&+&(-1)^{p(k-p)}u^{(2n+1)}_{p+2n+1,j-k+p+2m+1}u^{(2m+1)}_{k-p,j}),
\label{equationss1}
\end{eqnarray}
\begin{eqnarray}
D^{-}_n u^{(m)}_{k,j}&=&\sum^{n-1}_{p=0}
((-1)^{(p+n)m}v^{(n)}_{p,j}u^{(m)}_{k+p-n,j+p-n} \nonumber\\
&-&(-1)^{(p+n)(k+p-n)}v^{(n)}_{p,j-k-p+n+m}u^{(m)}_{k+p-n,j}),
\label{equationss2}
\end{eqnarray}
\begin{eqnarray}
D^{+}_n v^{(m)}_{k,j}&=&\sum^{n}_{p=0}
((-1)^{(p+n)m}u^{(n)}_{p,j}v^{(m)}_{k+p-n,j-p+n} \nonumber\\
&-&(-1)^{(p+n)(k+p-n)}u^{(n)}_{p,j+k+p-n-m}v^{(m)}_{k+p-n,j}),
\label{equationss3}
\end{eqnarray}
\begin{eqnarray}
D^{-}_n v^{(2m)}_{k,j}&=&\sum^{n-1}_{p=0}
(v^{(n)}_{p,j}v^{(2m)}_{k-p+n,j+p-n} \nonumber\\
&-&(-1)^{(p+n)(k-p+n)}v^{(n)}_{p,j+k-p+n-2m}v^{(2m)}_{k-p+n,j}),
\nonumber\\ D^{-}_{2n} v^{(2m+1)}_{k,j}&=&\sum^{2n-1}_{p=0}
((-1)^{p}v^{(2n)}_{p,j}v^{(2m+1)}_{k-p+2n,j+p-2n} \nonumber\\
&-&(-1)^{p(k-p)}v^{(2n)}_{p,j+k-p+2n-2m-1}v^{(2m+1)}_{k-p+2n,j}),
\nonumber\\ D^{-}_{2n+1} v^{(2m+1)}_{k,j}&=&\sum^{k}_{p=0}
((-1)^{p+1}v^{(2n+1)}_{p+2n+1,j}v^{(2m+1)}_{k-p,j+p} \nonumber\\
&+&(-1)^{p(k-p)}v^{(2n+1)}_{p+2n+1,j+k-p-2m-1}v^{(2m+1)}_{k-p,j})
\label{equationss4b}
\end{eqnarray}
(в правых частях все поля $\{u^{(m)}_{k,j},~v^{(m)}_{k,j}\}$ с $k
< 0$ должны быть положены равными нулю).

{\Large п}редставление {\Large л}акса \p{laxreprs1} генерирует неабелеву алгебру
потоков $N=(1|1)$ 2DTL иерархии
\begin{eqnarray}
[D^{+}_{n}~,~D^{-}_{l}\}=[D^{\pm}_{n}~,~D^{\pm}_{2l}]=0,\quad
\{D^{\pm}_{2n+1}~,~D^{\pm}_{2l+1}\}=2D^{\pm}_{2(n+l+1)},
\label{algebras1}
\end{eqnarray}
которая может быть реализована как
\begin{eqnarray}
D^{\pm}_{2n} ={\partial}^{\pm}_{2n}, \quad D^{\pm}_{2n+1}
={\partial}^{\pm}_{2n+1}+
\sum^{\infty}_{l=1}t^{\pm}_{2l-1}{\partial}^{\pm}_{2(k+l)},\quad
\partial^{\pm}_n:={\frac{\partial}{\partial t^{\pm}_n}},
\label{covder}
\end{eqnarray}
где $t^{\pm}_{2n}$ ($t^{\pm}_{2n+1}$) -- бозонные (фермионные)
эволюционные времена. {\Large н}епрерывный предел потоков
(\ref{equationss1}--\ref{equationss4b}) будет построен в параграфе
3.1 и ляжет в основу определения потоков бездисперсионной
$N=(1|1)$ 2DTL иерархии.

\subsection{{\Large б}озонные симметрии N=(1$|$1) 2DTL уравнения}

$\quad${\Large с}уперсимметричное $N=(1|1)$ 2DTL уравнение
\begin{eqnarray}
D^{+}_1D^{-}_1 \ln v_{0,j}= v_{0,j+1} - v_{0,j-1} \label{todabs}
\end{eqnarray}
принадлежит системе уравнений
(\ref{equationss1}--\ref{equationss4b}). Оно вытекает из уравнения
\p{equationss2} при $\{n=m=k=1\}$
\begin{eqnarray}
D^{-}_1 u_{1,j} = - v_{0,j} - v_{0,j+1} \label{todabs1}
\end{eqnarray}
и уравнения \p{equationss3} при $\{n=m=1, k=0\}$
\begin{eqnarray}
D^{+}_1 v_{0,j}= v_{0,j}(u_{1,j} - u_{1,j-1}) \label{todabs2}
\end{eqnarray}
после исключения поля $u_{1,j}$. Его бозонные симметрии
$D^{\pm}_{2n} v_{0,j}$
\begin{eqnarray}
\{D^{+}_1,D^{\pm}_{2n}\}= \{D^{-}_1,D^{\pm}_{2n}\}=0
\label{algebra2new}
\end{eqnarray}
были описаны в работах \cite{ls1,ols1,vgks} в терминах следующей
итеррационной процедуры:
\begin{eqnarray}
D^{\pm}_{2n}
v_{0,j}&=&v_{0,j}(u^{(2n)\pm}_{2n,j}-u^{(2n)\pm}_{2n,j-1}), \quad
u^{(2n)\pm}_{0,j}=1, \nonumber\\ \pm D^{\mp}_1
u^{(2n)\pm}_{k,j}&=&v_{0,j}u^{(2n)\pm}_{k-1,j-1} +
(-1)^{k}v_{0,j-k+2n+1}u^{(2n)\pm}_{k-1,j}, \label{flowss+-bos}
\end{eqnarray}
где функции $u^{(n)\pm}_{k,j}$ связаны с исходными функционалами
$\{u^{(n)}_{k,j}, v^{(n)}_{k,j}\}$ следующим образом:
\begin{eqnarray}
u^{(n)+}_{k,j}= u^{(n)}_{k,j}, \quad u^{(n)-}_{k,j}=
\frac{v^{(n)}_{k,-j-1}}{{\sum}^{n-k}_{m=1} v_{0,k+m-n-j-1}}
\label{relation}
\end{eqnarray}
(детали см. в \cite{vgks}).

{\Large н}епрерывный предел $N=(1|1)$ 2DTL уравнения \p{todabs} и его
симметрий $D^{\pm}_{2n} v_{0,j}$ \p{flowss+-bos} будет получен в
параграфе 3.2.

\section{{\Large б}ездисперсионная N=(1$|$1) 2DTL иерархия}

$\quad$ {\Large в} этом параграфе строится непрерывный (квазиклассический)
по шагу решетки предел $N=(1|1)$ 2DTL иерархии ---
бездисперсионная $N=(1|1)$ 2DTL иерархия, и конструируется
соответствующее ему представление {\Large л}акса.

\subsection{Квазиклассический предел}

$\quad$ {\Large п}отоки (\ref{equationss1}--\ref{equationss4b}) $N=(1|1)$
2DTL иерархии, введеные в предыдущем параграфе, не содержат явную
зависимость от размерных постоянных, а решетка с безразмерной
координатой $j$ ($j\in {\bZ})$ имеет единичный шаг. Для изучения
непрерывного предела введем явно длину шага решетки. {\Large п}оскольку
этот параметр в дальнейшем будет играть роль постоянной {\Large п}ланка, мы
будем обозначать его как ${\hbar}$. Таким образом, вместо $j$
возникает комбинация
\begin{eqnarray}
{\hbar} j \equiv s, \label{coord1}
\end{eqnarray}
и все решеточные поля начинают зависеть от параметра ${\hbar}$.
Тогда, непрерывный (квазиклассический) предел может быть определен
как
\begin{eqnarray}
{\hbar}\rightarrow 0, \quad s = {\lim}_{{\hbar}\rightarrow 0, ~j
>> 1}({\hbar} j), \label{displimit1}
\end{eqnarray}
а $s$ выступает в роли непрерывной "решеточной" координаты.

Для того, чтобы потоки (\ref{equationss1}--\ref{equationss4b})
были нетривиальными и регулярными в пределе \p{displimit1},
необходимо дополнительно осуществить некоторые масштабные
преобразования зависимых и независимых переменных, принадлежащих
системе. Так, мы постулируем следующие правила перехода к новым
эволюционным временам
\begin{eqnarray}
t^{\pm}_{2n+1} \rightarrow \frac{1}{\sqrt{\hbar}} t^{\pm}_{2n+1},
~~ t^{\pm}_{2n} \rightarrow \frac{1}{{\hbar}} t^{\pm}_{2n}
~\Leftrightarrow ~ D^{\pm}_{2n+1}  \rightarrow \sqrt{\hbar}
D^{\pm}_{2n+1}, ~~ D^{\pm}_{2n}  \rightarrow {\hbar}
D^{\pm}_{2n}~~~~~ \label{displimit2}
\end{eqnarray}
и полям иерархии
\begin{eqnarray}
&&u_{2k,j} \quad ~\rightarrow ~~~~u_{2k}({\hbar} j), \quad
\quad ~~~v_{2k,j} ~\quad \rightarrow ~~~~v_{2k}({\hbar} j),\nonumber\\
&&u_{2k+1,j}~\rightarrow ~ \frac{1}{\sqrt{\hbar}}~u_{2k+1}({\hbar}
j), \quad ~~v_{2k+1,j}~~\rightarrow ~
\frac{1}{\sqrt{\hbar}}~v_{2k+1}({\hbar} j), \label{displimit3}
\end{eqnarray}
полагая их несингулярными при ${\hbar}=0$.

Теперь можно установить два нетривиальные ключевые свойства
квазиклассического предела (\ref{displimit1}--\ref{displimit3}),
которые имеют важное значение для дальнейшего рассмотрения и
проверяются прямыми вычислениями.

{\Large п}ервое свойство состоит в том, что новые композитные поля,
определенные по следующим правилам
\begin{eqnarray}
&&u^{(m)}_{2k,j} \quad ~\rightarrow ~~~~u^{(m)}_{2k}({\hbar} j),
\quad \quad ~v^{(m)}_{2k,j} \quad \rightarrow
~~~~~v^{(m)}_{2k}({\hbar} j), \nonumber\\
&&u^{(2m+1)}_{2k+1,j}\rightarrow ~\frac{1}{\sqrt{\hbar}}~
u^{(2m+1)}_{2k+1}({\hbar}j), \quad v^{(2m+1)}_{2k+1,j}\rightarrow
~ \frac{1}{\sqrt{\hbar}}~v^{(2m+1)}_{2k+1}({\hbar} j),\nonumber\\
&&u^{(2m)}_{2k+1,j}~\rightarrow ~\sqrt{\hbar}~
u^{(2m)}_{2k+1}({\hbar} j), \quad \quad
v^{(2m)}_{2k+1,j}~\rightarrow ~\sqrt{\hbar}~
v^{(2m)}_{2k+1}({\hbar}j), \label{displimit4}
\end{eqnarray}
регулярны в квазиклассическом пределе.

{\Large и}з (\ref{displimit1}--\ref{displimit4}) и очевидных тождеств
\begin{eqnarray}
(L^{{\alpha}})^{2(m+1)}_{*}:= (L^{{\alpha}})^{2}_{*}
(L^{{\alpha}})^{2m}_{*}, \quad (L^{{\alpha}})^{(2m+1)}_{*}:=
L^{{\alpha}} (L^{{\alpha}})^{2m}_{*}, \label{ident1}
\end{eqnarray}
следующих из \p{rule2}, можно получить, например, важные рекуррентные
соотношения для лидирующих членов функционалов
$u^{(m)}_{k}\equiv u^{(m)}_{k}(s)$
\begin{eqnarray}
&& u^{(2(l+1))}_{2k}
=\sum^{k}_{n=0}u^{(2)}_{2n}u^{(2l)}_{2(k-n)},\quad
u^{(2(l+1))}_{2k+1}=\sum^{2k+1}_{n=0}u^{(2)}_{n}u^{(2l)}_{2k-n+1},
\nonumber\\
&& u^{(2l+1)}_{2k} ~=\sum^{2k}_{n=0}u_{n}u^{(2l)}_{2k-n},\quad
\quad ~~u^{(2l+1)}_{2k+1}=\sum^{k}_{n=0}u_{2n+1}u^{(2l)}_{2(k-n)},
\label{rec1}
\end{eqnarray}
\begin{eqnarray}
&& u^{(2)}_{2k} =\sum^{k}_{n=0}u_{2n}u_{2(k-n)} +2\sum^{k-1}_{n=0}
(k-n-1)u_{2(k-n)-1}{\partial}_su_{2n+1}, \nonumber\\
&& u^{(2)}_{2k+1} =\sum^{k}_{n=0}\Big[(1-2n)u_{2n}{\partial}_s
u_{2(k-n)+1} +2(k-n)u_{2(k-n)+1} {\partial}_su_{2n}\Big],
\label{rec2}
\end{eqnarray}
где ${\partial}_s:= \frac{{\partial}}{{\partial s}}$.

{\Large в}торое свойство сводится к тому, что в квазиклассическом пределе
(\ref{displimit1}--\ref{displimit4}) потоки
(\ref{equationss1}--\ref{equationss4b}) $N=(1|1)$ 2DTL иерархии
нетривиальны и регулярны. Так, явные выражения для их лидирующих
членов следующие:
\begin{eqnarray}
при \quad \{m=2l, ~&k&=2r\}, \quad или \quad \{m=2l+1, ~k=2r+1\} \nonumber\\
D^{-}_{2n+1} u^{(m)}_{k}&=&\sum^{n-1}_{p=0}
\Bigl[2(p-n)v^{(2n+1)}_{2p+1}{\partial}_s u^{(m)}_{k+2(p-n)}\nonumber\\
&+&(k-m+2(p-n))({\partial}_sv^{(2n+1)}_{2p+1})u^{(m)}_{k+2(p-n)}\Bigr]
\nonumber\\
&+&2(-1)^k\sum^{n}_{p=0}v^{(2n+1)}_{2p}u^{(m)}_{k+2(p-n)-1},
\nonumber\\
D^{-}_{2n} u^{(m)}_{k}&=&\sum^{n-1}_{p=0}
\Bigl[2(p-n)v^{(2n)}_{2p}{\partial}_s u^{(m)}_{k+2(p-n)}\nonumber\\
&+&(k-m+2(p-n))({\partial}_sv^{(2n)}_{2p})u^{(m)}_{k+2(p-n)}
\nonumber\\
&+&2(-1)^kv^{(2n)}_{2p+1}u^{(m)}_{k+2(p-n)+1}\Bigr], \nonumber\\
D^{+}_{2n} u^{(m)}_{k}&=&\sum^{n}_{p=0}
\Bigl[2(n-p)u^{(2n)}_{2p}{\partial}_s u^{(m)}_{k+2(n-p)}\nonumber\\
&+&(k-m+2(n-p))({\partial}_su^{(2n)}_{2p})u^{(m)}_{k+2(n-p)}\Bigr]
\nonumber\\&+& 2(-1)^k \sum^{n-1}_{p=0}
u^{(2n)}_{2p+1}u^{(m)}_{k+2(n-p)-1}, \nonumber\\
D^{+}_{2n+1} v^{(m)}_{k}&=&\sum^{n}_{p=0}
\Bigl[2(n-p)u^{(2n+1)}_{2p+1}{\partial}_s v^{(m)}_{k+2(p-n)}\nonumber\\
&-&(k-m+2(p-n))({\partial}_su^{(2n+1)}_{2p+1})v^{(m)}_{k+2(p-n)} \nonumber\\
&+& 2(-1)^k u^{(2n+1)}_{2p}v^{(m)}_{k-1+2(p-n)}\Bigr], \nonumber\\
D^{+}_{2n} v^{(m)}_{k}&=&\sum^{n}_{p=0}
\Bigl[2(n-p)u^{(2n)}_{2p}{\partial}_s v^{(m)}_{k+2(p-n)}\nonumber\\
&-&(k-m+2(p-n))({\partial}_su^{(2n)}_{2p})v^{(m)}_{k+2(p-n)}\Bigr]
\nonumber\\&+& 2(-1)^k \sum^{n-1}_{p=0}
u^{(2n)}_{2p+1}v^{(m)}_{k+1+2(p-n)}, \nonumber\\
D^{-}_{2n} v^{(m)}_{k}&=&\sum^{n-1}_{p=0}
\Bigl[2(p-n)v^{(2n)}_{2p}{\partial}_s v^{(m)}_{k+2(n-p)}\nonumber\\
&-&(k-m+2(n-p))({\partial}_sv^{(2n)}_{2p})v^{(m)}_{k+2(n-p)}\nonumber\\
&+&2(-1)^kv^{(2n)}_{2p+1}v^{(m)}_{k+2(n-p)-1}\Bigr],
\label{equ1+}
\end{eqnarray}
и
\begin{eqnarray}
при \quad \{m=2l, ~&k&=2r+1\}, \quad или \quad \{m=2l+1, ~k=2r\} \nonumber\\
D^{-}_n u^{(m)}_{k}&=&\sum^{n-1}_{p=0}
(-1)^{m(p+n)}\Bigl[(p-n)v^{(n)}_{p}{\partial}_s u^{(m)}_{k+p-n}\nonumber\\
&+&(k+p-n-m)({\partial}_sv^{(n)}_{p})u^{(m)}_{k+p-n}\Bigr],
\nonumber\\ D^{+}_{2n} u^{(m)}_{k}&=&\sum^{2n}_{p=0}
(-1)^{mp}\Bigl[(2n-p)u^{(2n)}_{p}{\partial}_s u^{(m)}_{k-p+2n}\nonumber\\
&+&(k-p+2n-m)({\partial}_su^{(2n)}_{p})u^{(m)}_{k-p+2n}\Bigr],
\nonumber\\ D^{+}_{n} v^{(m)}_{k}&=&\sum^{n}_{p=0}
(-1)^{m(p+n)}\Bigl[(n-p)u^{(n)}_{p}{\partial}_s v^{(m)}_{k+p-n}\nonumber\\
&-&(k+p-n-m)({\partial}_su^{(n)}_{p})v^{(m)}_{k+p-n}\Bigr],
\nonumber\\
D^{-}_{2n} v^{(m)}_{k}&=&\sum^{2n-1}_{p=0}
(-1)^{mp}\Bigl[(p-2n)v^{(2n)}_{p}{\partial}_s v^{(m)}_{k-p+2n}\nonumber\\
&-&(k-p+2n-m)({\partial}_sv^{(2n)}_{p})v^{(m)}_{k-p+2n}\Bigr],
\label{equ2+}
\end{eqnarray}
а также
\begin{eqnarray}
D^{+}_{2n+1} u^{(2m)}_{2k}&=&2\sum^{n}_{p=0}
\Bigl[(n-p)u^{(2n+1)}_{2p+1}{\partial}_s u^{(2m)}_{2(k-p+n)}\nonumber\\
&+&(k-p+n-m)({\partial}_su^{(2n+1)}_{2p+1})u^{(2m)}_{2(k-p+n)}
\nonumber\\ &+& u^{(2n+1)}_{2p}u^{(2m)}_{2(k-p+n)+1}\Bigr],
\nonumber\\ D^{+}_{2n+1} u^{(2m+1)}_{2k+1}&=&2\sum^{k}_{p=1}
\Bigl[pu^{(2n+1)}_{2(p+n)+1}{\partial}_s u^{(2m+1)}_{2(k-p)+1}\nonumber\\
&+&(m+p-k)({\partial}_su^{(2n+1)}_{2(p+n)+1})u^{(2m+1)}_{2(k-p)+1}\Bigr]
\nonumber\\ &+& 2\sum^{k}_{p=0}
u^{(2n+1)}_{2(p+n+1)}u^{(2m+1)}_{2(k-p)},\nonumber\\
D^{+}_{2n+1} u^{(2m+1)}_{2k}&=&\sum^{2k}_{p=1}
(-1)^p\Bigl[pu^{(2n+1)}_{p+2n+1}{\partial}_s u^{(2m+1)}_{2k-p}\nonumber\\
&+&(2(m-k)+p+1)({\partial}_su^{(2n+1)}_{p+2n+1})u^{(2m+1)}_{2k-p}\Bigr],
\nonumber\\ D^{+}_{2n+1} u^{(2m)}_{2k+1}&=&\sum^{2n+1}_{p=0}
\Bigl[(2n-p+1)u^{(2n+1)}_{p}{\partial}_s u^{(2m)}_{2(k+n+1)-p}\nonumber\\
&+&(2(k+n-m+1)-p)({\partial}_su^{(2n+1)}_{p})u^{(2m)}_{2(k+n+1)-p}\Bigr],
\nonumber\\D^{-}_{2n+1} v^{(2m)}_{2k}&=&2\sum^{n-1}_{p=0}
\Bigl[(p-n)v^{(2n+1)}_{2p+1}{\partial}_s v^{(2m)}_{2(k-p+n)}\nonumber\\
&-&(k-p+n-m)({\partial}_sv^{(2n+1)}_{2p+1})v^{(2m)}_{2(k-p+n)}\Bigr]
\nonumber\\ &+&2\sum^{n}_{p=0}
v^{(2n+1)}_{2p}v^{(2m)}_{2(k-p+n)+1}, \nonumber\\D^{-}_{2n+1}
v^{(2m+1)}_{2k+1}&=&2\sum^{k}_{p=0}
\Bigl[-pv^{(2n+1)}_{2(p+n)+1}{\partial}_s v^{(2m+1)}_{2(k-p)+1}\nonumber\\
&+&(k-p-m){\partial}_sv^{(2n+1)}_{2(p+n)+1})v^{(2m+1)}_{2(k-p)+1}
\nonumber\\ &+& v^{(2n+1)}_{2(p+n+1)}v^{(2m+1)}_{2(k-p)}\Bigr],
\nonumber\\D^{-}_{2n+1} v^{(2m+1)}_{2k}&=&\sum^{k}_{p=0}
(-1)^p\Bigl[-pv^{(2n+1)}_{p+2n+1}{\partial}_s v^{(2m+1)}_{2k-p}\nonumber\\
&-&(2(m-k)+p+1)({\partial}_sv^{(2n+1)}_{p+2n+1})v^{(2m+1)}_{2k-p}\Bigr],
\nonumber\\D^{-}_{2n+1} v^{(2m)}_{2k+1}&=&\sum^{2n}_{p=0}
\Bigl[-(2n-p+1)v^{(2n+1)}_{p}{\partial}_s v^{(2m)}_{2(k+n+1)-p}\nonumber\\
&-&(2(k+n-m+1)-p)({\partial}_sv^{(2n+1)}_{p})v^{(2m)}_{2(k+n+1)-p}\Bigr],
\label{eqvv6-}
\end{eqnarray}
где в правых частях все поля $\{u^{(m)}_{k},~v^{(m)}_{k}\}$ с $k < 0$
должны быть положены равными нулю. {\Large п}олученные таким способом потоки
(\ref{equ1+}--\ref{eqvv6-}) мы называем бездисперсионной $N=(1|1)$
2DTL иерархией. {\Large п}редставление {\Large л}акса для них будет построено в
параграфе 3.3.

\subsection{{\Large б}ездисперсионное N=(1$|$1) 2DTL уравнение и его бозонные
симметрии}

$\quad$ {\Large б}ездисперсионные пределы $N=(1|1)$ 2DTL уравнения
\p{todabs} и его бозонных симметрий (\ref{flowss+-bos}) могут быть
получены без труда, если использовать соотношения
(\ref{relation}--\ref{displimit4}). {\Large п}олучаем, сответственно:
\begin{eqnarray}
D^{+}_1D^{-}_1 \ln v_{0}= 2 {\partial}_s v_{0}
\label{disptoda}
\end{eqnarray}
и
\begin{eqnarray}
D^{\pm}_{2n} v_{0}&=&v_{0}{\partial}_s
u^{(2n)\pm}_{2n}, \quad u^{(2n)\pm}_{0,j}=1, \nonumber\\
\mp D^{\mp}_1 u^{(2n)\pm}_{2k+1}&=&v_{0}{\partial}_s
u^{(2n)\pm}_{2k} + 2(n-k)({\partial}_sv_{0})u^{(2n)\pm}_{2k},
\nonumber\\ \pm D^{\mp}_1
u^{(2n)\pm}_{2k}&=&2v_{0}u^{(2n)\pm}_{2k-1}.
\label{disp-flowss+-bos}
\end{eqnarray}

{\Large и}сключая $u^{(2n)\pm}_{2k+1}$ из системы уравнений
\p{disp-flowss+-bos}, окончательно получаем следующую рекуррентную
систему уравнений для генерации бозонных симметрий
бездисперсионного $N=(1|1)$ 2DTL уравнения \p{disptoda}
\begin{eqnarray}
D^{\pm}_{2n} v_{0}&=&v_{0}{\partial}_s
u^{(2n)\pm}_{2n}, \quad u^{(2n)\pm}_{0,j}=1, \nonumber\\
-D^{\mp}_1 u^{(2n)\pm}_{2k}&=&2v_{0} {(D^{\mp}_1)}^{-1} \Bigl[
v_{0}{\partial}_s u^{(2n)\pm}_{2(k-1)} +
2(n-k+1)({\partial}_sv_{0})u^{(2n)\pm}_{2(k-1)}\Bigr].
\label{disp-flowss+-bos1}
\end{eqnarray}
Отметим, что симметрии бозонного бездисперсионного 2DTL уравнения
были найдены в работе \cite{fl} (см. также работу \cite{hp}) путем
решения соответствующего, достаточно сложного уравнения симметрии.

\subsection{{\Large п}редставление {\Large л}акса}
%\subsection{Суперскобки Пуассона}

$\quad$ {\Large в} параграфе 3.1 были построены потоки
(\ref{equ1+}--\ref{eqvv6-}) бездисперсионной $N=(1|1)$ 2DTL
иерархии. {\Large н}ашей задачей сейчас будет нахождение соответствующего
им представления {\Large л}акса. {\Large п}оскольку представление {\Large л}акса
\p{laxreprs1} $N=(1|1)$ 2DTL иерархии выражается в терминах
обобщенной градуированной скобки \p{brack}, из предыдущего опыта
естественно ожидать, что для получения его квазиклассического
предела, соответствующего бездисперсионной $N=(1|1)$ 2DTL
иерархии, достаточно заменить в нем эту скобку на определенную
суперскобку {\Large п}уассона, а операторы - на их символы, заданные на
соответствующем фазовом пространстве.

{\Large н}ашей ближайшей задачей будет моделирование фазового пространства
и суперскобки {\Large п}уассона, исходя из некоторых свойств, которыми они
должны быть наделены. Так, вспоминая, что операторы {\Large л}акса
$N=(1|1)$ 2DTL иерархии, будучи задаными на пространстве
операторов, градуированном двумя разными $Z_2$--градуировками
(\ref{grad1}) и (\ref{grad2}), обладают только одной диагональной
$Z_2$--градуировкой \p{gradop}, можно полагать, что фазовое
пространство унаследует эти свойства. {\Large п}ринимая во внимание
$Z_2$--градуировку $d_{e^{{\partial}}}=1, ~d_{e^{2{\partial}}}=0$
(\ref{grad2}) оператора решеточного сдвига $e^{l{\partial}}$, мы
предполагаем, что на фазовом пространстве ему соответствуют две
координаты, а именно, грассманова координата $\pi$ ($\pi^2=0$) и
бозонная координата $p$, со следующими правилами соответствия:
\begin{eqnarray}
e^{{\partial}}~\rightarrow ~ \frac{1}{{\sqrt\hbar}}~\pi, \quad
 e^{2{\partial}}~\rightarrow  ~p . \label{displax1}
\end{eqnarray}
{\Large н}есмотря на грассмановость координаты $\pi$, она должна обладать
{\it "нетипичным"} для грассмановых координат свойством коммутировать не
только с бозонными, но и фермионными полями иерархии, что следует
из непрерывного предела соотношения \p{rule1}.

Очевидно, что, кроме координат $\pi$ и $p$, фазовое пространство
должно также включать непрерывную "решеточную" координату $s$
\p{coord1}. {\Large в}озникающее в результате фазовое суперпространство
$\{\pi, ~p,~ s\}$ бездисперсионной $N=(1|1)$ 2DTL иерархии
содержит фазовое пространство $\{p,~ s\}$ бездисперсионной 2DTL
иерархии как подпространство.

{\Large п}осле того, как координаты фазового пространства установлены, мы
должны построить для них суперскобки {\Large п}уассона. {\Large в}споминая, что
суперскобки {\Large п}уассона должны находиться в соответствии с
cупералгеброй $Z_2$--градуированных операторов
$\{\sqrt{\hbar}e^{{\partial}}, ~e^{2{\partial}},~ {\hbar} j\}$,
которым отвечают координаты фазового суперпространства $\{\pi,
~p,~ s\}$, эти суперскобки могут быть сравнительно легко
сконструированы. {\Large п}риведем здесь суперскобки {\Large п}уассона между двумя
произвольными функциями ${\bF}_{1,2}\equiv {\bF}_{1,2}(\pi,p,s)$
на фазовом пространстве, полученные способом, изложенным выше:
\begin{eqnarray}
\Bigl\{{\bF}_1,{\bF}_2\Bigr\}&=& 2p~\Bigl(\frac{\partial
{\bF}_1}{\partial p} ~\frac{\partial {\bF}_2}{\partial s}-
\frac{\partial {\bF}_1}{\partial s} ~\frac{\partial
{\bF}_2}{\partial p}+ \frac{\partial {\bF}_1}{\partial \pi}~
\frac{\partial {\bF}_2}{\partial \pi}\Bigr)\nonumber\\
&+&\pi ~\Bigl(\frac{\partial {\bF}_1}{\partial \pi}~
\frac{\partial {\bF}_2}{\partial s}- \frac{\partial
{\bF}_1}{\partial s}~ \frac{\partial {\bF}_2}{\partial \pi}\bigr).
\label{poissbrack}
\end{eqnarray}

{\Large з}десь уместно отметить, что после перехода к новому базису
$\{{\widetilde s}, \widetilde p,\widetilde \pi\}$ фазового
пространства, задаваемому формулами
\begin{eqnarray}
{\widetilde s}&:=&\frac{s}{2}, \quad \widetilde p := \ln p, \quad
{\widetilde \pi} := \frac{\pi}{\sqrt{2p}}, \nonumber\\
s&:=&2{\widetilde s}, \quad p := e^{\widetilde p}, \quad \pi :=
\sqrt{2}{\widetilde \pi} e^{\frac{\widetilde p}{2}},
 \label{poissbrack2}
\end{eqnarray}
суперскобки {\Large п}уассона \p{poissbrack} принимают вид
\begin{eqnarray}
\Bigl\{{\bF}_1,{\bF}_2\Bigr\}&=& \frac{\partial {\bF}_1}{\partial
\widetilde p} ~\frac{\partial {\bF}_2}{\partial \widetilde s}-
\frac{\partial {\bF}_1}{\partial \widetilde s} ~\frac{\partial
{\bF}_2}{\partial \widetilde p}+ \frac{\partial {\bF}_1}{\partial
\widetilde \pi} ~\frac{\partial {\bF}_2}{\partial \widetilde \pi},
\label{poissbrack1}
\end{eqnarray}
что соответствует канонической ортосимплектической структуре
фазового суперпространства.

Теперь мы перейдем к следующему этапу получения представления
{\Large л}акса бездисперсионной $N=(1|1)$ 2DTL иерархии и приведем
эвристические формулы определения и построения символов ${\cal
L}^{{\pm}}$ и $({\cal L}^{{\pm}})^{2m}_{*}$ операторов {\Large л}акса
$L^{\pm}$ \p{laxs1} и композитных операторов $(L^{\pm})^{n}_{*}$
(\ref{rule2}--\ref{laxs2})
\begin{eqnarray}
L^{{\pm}}~ &\rightarrow & ~ \frac{1}{{\sqrt\hbar}}~{\cal
L}^{{\pm}},  \nonumber\\
{\cal L}^{+}=\sum^{\infty}_{k=0} (u_{2k+1}+u_{2k}\pi)p^{-k},
&\quad &{\cal L}^{-}=\sum^{\infty}_{k=0}
(v_{2k-1}+v_{2k}\pi)p^{k-1}
\label{displaxs1}
\end{eqnarray}
и
\begin{eqnarray}
&&(L^{{\pm}})^{2m}_{*}~ \rightarrow ~({\cal L}^{{\pm}})^{2m}_{*},
\quad (L^{{\pm}})^{2m+1}_{*}~ \rightarrow ~
\frac{1}{{\sqrt\hbar}}~({\cal L}^{{\pm}})^{2m+1}_{*}, \nonumber\\
&&\quad \quad \quad ({\cal L}^{+})^{2m}_{*}~~~:=~
\sum^{\infty}_{k=0} (u^{(2m)}_{2k}+u^{(2m)}_{2k-1}\pi)p^{m-k}, \nonumber\\
&&\quad \quad \quad ({\cal L}^{-})^{2m}_{*}~~~:=~
\sum^{\infty}_{k=0} (v^{(2m)}_{2k}+v^{(2m)}_{2k+1}\pi)p^{k-m}, \nonumber\\
&&\quad \quad \quad ({\cal L}^{+})^{2m+1}_{*}:=~
\sum^{\infty}_{k=0}
(u^{(2m+1)}_{2k+1}+u^{(2m+1)}_{2k}\pi)p^{m-k},\nonumber\\
&&\quad \quad \quad ({\cal L}^{-})^{2m+1}_{*}:=~
\sum^{\infty}_{k=0}
(v^{(2m+1)}_{2k-1}+v^{(2m+1)}_{2k}\pi)p^{k-m-1}, \label{diplaxs2}
\end{eqnarray}
соответственно. {\Large п}о определению, все поля
$\{u^{(m)}_{k},~v^{(m)}_{k}\}$ с $k < 0$ должны быть положены
равными нулю. {\Large п}ри получении этих выражений мы использовали
подстановки \p{displimit4} и \p{displax1}.
{\Large з}аметим, что эти символы в общем случае некоммутативны
\begin{eqnarray}
({\cal L}^{{\alpha}})^{k}_{*} ({\cal L}^{{\beta}})^{m}_{*}=
(-1)^{km}(({\cal L}^{{\beta}})^{m}_{*})^{*(k)} (({\cal
L}^{{\alpha}})^{k}_{*})^{*(m)}, \quad {\alpha},{\beta} =+,-,
\label{dispprop3}
\end{eqnarray}
что связано с {\it "нетипичным"} свойством грассмановой координаты $\pi$,
отмеченным выше (см. параграф после \p{displax1}).

{\Large и}, наконец, заменяя операторы {\Large л}акса на их символы
(\ref{displaxs1}--\ref{diplaxs2}), а обобщенную градуированную
скобку (\ref{brack}) на суперскобку {\Large п}уассона \p{poissbrack} в
представлениях {\Large л}акса (\ref{laxreprs1}) и \p{laxreprs2} для
$N=(1|1)$ 2DTL иерархии и в определяющих его соотношениях
(\ref{rule2}) по правилу
\begin{eqnarray}
\Bigl[~ \ldots ~, ~\ldots ~\Bigr\} ~\rightarrow  ~
{\hbar}~\Bigl\{~\ldots~,~\ldots~\Bigr\}, \label{dispbrac}
\end{eqnarray}
осуществляя затем подстановку \p{displimit2} и переходя к пределу
\p{displimit1}, мы получаем выражения
%\newpage
\begin{eqnarray}
D^{\pm}_n {\cal L}^{{\alpha}}&=& \mp {\alpha} (-1)^n \Bigr\{
{((({\cal L}^{\pm})^{n}_{*})}_{-{\alpha} })^{*}, {\cal
L}^{{\alpha}}\Bigl\}, \quad n \in {\bN}, \quad {\alpha} =+,-,
\label{displaxreprs1}
\end{eqnarray}
\begin{eqnarray}
D^{\pm}_n ({\cal L}^{{\alpha}})^{m}_{*}&=& \mp {\alpha} (-1)^{nm}
\Bigr\{ {((({\cal L}^{\pm})^{n}_{*})}_{-{\alpha} })^{*(m)}, ({\cal
L}^{{\alpha}})^{m}_{*}\Bigl\},\quad n,m \in {\bN},
\label{displaxreprs2}
\end{eqnarray}
\begin{eqnarray}
({\cal L}^{{\alpha}})^{2m}_{*}:= \Bigl(~\frac{1}{2}~\Bigl\{({\cal
L}^{{\alpha}})^{*},{\cal L}^{{\alpha}}\Bigr\}~\Bigr)^{m},\quad
({\cal L}^{{\alpha}})^{2m+1}_{*}:={\cal L}^{{\alpha}}~({\cal
L}^{{\alpha}})^{2m}_{*} \label{disprule2}
\end{eqnarray}
для бездисперсионной $N=(1|1)$ 2DTL иерархии. {\Large п}рямые вычисления
подтверждают, что полученные соотношения
(\ref{displaxreprs2}--\ref{disprule2}) правильно воспроизводят
бездисперсионные потоки (\ref{rec1}--\ref{eqvv6-}), т.е. являются
{\it искомым представлением {\Large л}акса для бездисперсионной $N=(1|1)$
2DTL иерархии.}

\section{{\Large з}аключение}

$\quad$ {\Large в} данной работе мы предложили скобочную операцию
(\ref{brack}) со свойствами (\ref{prop1}--\ref{jacobi}) на
пространстве градуированных операторов с инволюцией, обобщающую
градуированный коммутатор супералгебр. {\Large з}атем мы получили новую
форму представления {\Large л}акса (\ref{laxreprs1}--\ref{rule2}) и
\p{laxreprs2} для двумерной $N=(1|1)$ суперсимметричной решеточной
иерархии Тоды в терминах обобщенной градуированной скобки. Далее
мы использовали это представление при построении
квазиклассического предела (\ref{displimit1}--\ref{displimit4})
этой иерархии --- бездисперсионной $N=(1|1)$ иерархии Тоды
(\ref{equ1+}--\ref{eqvv6-}), и его представления {\Large л}акса
(\ref{displaxs1}--\ref{diplaxs2}),
(\ref{displaxreprs1}--\ref{disprule2}) на градуированном фазовом
суперпространстве со скобкой {\Large п}уассона \p{poissbrack}. {\Large и}, наконец,
попутно мы установили бозонные симметрии \p{disp-flowss+-bos1}
бездисперсионного $N=(1|1)$ суперсимметричного уравнения Тоды
\p{disptoda}.

{}~

{\it Один из нас ({\Large в.к.}) впервые познакомился с {\Large а.а. л}огуновым, еще
будучи студентом II курса {\Large ф}изического факультета {\Large мгу} (1956), и на
протяжении прошедших десятилетий ощущал с его стороны огромную
поддержку как в делах, связанных с наукой, так и в жизни. {\Large б}удьте
здоровы, дорогой {\Large а}натолий {\Large а}лексеевич!}

\end{document}

\def\l#1#2{\raisebox{.0ex}{$\displaystyle
  \mathop{#1}^{{\scriptstyle #2}\rightarrow}$}}
\def\r#1#2{\raisebox{.0ex}{$\displaystyle
\mathop{#1}^{\leftarrow {\scriptstyle #2}}$}}

\newcommand{\p}[1]{(\ref{#1})}

%%%%%%%%%%%%
\newcommand{\sect}[1]{\setcounter{equation}{0}\section{#1}}
%%%%%%%%%%%%
%\renewcommand{\theequation}{\thesection.\arabic{equation}}

\makeatletter
\def\eqnarray{\stepcounter{equation}\let\@currentlabel=\theequation
\global\@eqnswtrue
\global\@eqcnt\z@\tabskip\@centering\let\\=\@eqncr
$$\halign to \displaywidth\bgroup\@eqnsel\hskip\@centering
  $\displaystyle\tabskip\z@{##}$&\global\@eqcnt\@ne
  \hfil$\displaystyle{\hbox{}##\hbox{}}$\hfil
  &\global\@eqcnt\tw@ $\displaystyle\tabskip\z@
  {##}$\hfil\tabskip\@centering&\llap{##}\tabskip\z@\cr}
%\@addtoreset{equation}{section}
%  \def\theequation{\thesection.\arabic{equation}}
\makeatother

\renewcommand{\thefootnote}{\fnsymbol{footnote}}
%\newpage
%\setcounter{page}{0}
%\pagestyle{empty}
%\begin{flushright}
%{Февраль 2002}\\
%{JINR E2-???-2002}\\
%{nlin.SI/0206044}
%\end{flushright}

%\vfill

%%%%%%%%%%%%%%%%%%%%%%%%%%

\begin{center}
{\sl Theoretical and Mathematical Physics\footnote{Translated from 
Teoreticheskaya i Matematicheskaya Fizika
{\bf 132} (2002) 222.}\\
{\bf 132} (2002) 1080, nlin.SI/0206044}
\end{center}

\begin{center}
{\LARGE {\bf N=(1$|$1) supersymmetric }}\\[0.3cm]
{\LARGE {\bf dispersionless}}\\[0.3cm]
{\LARGE {\bf Toda lattice hierarchy}}\\[0.5cm]

{}~

{\large V.G. Kadyshevsky$^{(a)}$ and A.S. Sorin$^{(b)}$}

{}~\\
\quad \\

{{Bogoliubov Laboratory of Theoretical Physics,}}\\
{{Joint Institute for Nuclear Research,}}\\
{\em 141980 Dubna, Moscow Region, Russia}~\quad\\
\em {~$~^{(a)}$ Email: kadyshev@jinr.ru}\\
\em {~$~^{(b)}$ Email: sorin@thsun1.jinr.ru}\\

\end{center}

%\vfill

{}~

\centerline{{\bf Abstract}}
\noindent
Generalizing the graded commutator in superalgebras, we propose a new 
bracket
operation on the space of graded operators with an involution. We study
properties of this operation and show that the Lax representation of the
two-dimensional $N{=}(1|1)$ supersymmetric Toda lattice hierarchy can be
realized via the generalized bracket operation; this is important in
constructing the semiclassical (continuum) limit of this 
hierarchy. We construct the continuum limit of the $N{=}(1|1)$ Toda 
lattice hierarchy, the dispersionless $N{=}(1|1)$ Toda hierarchy. In this 
limit, we obtain the Lax representation, with the generalized graded 
bracket becoming the corresponding Poisson bracket on the graded phase 
superspace. We find bosonic symmetries of the dispersionless $N{=}(1|1)$ 
supersymmetric Toda equation.   
\vskip 2mm

\begin{center} 
\noindent{\sl Dedicated to the $75$th birthday of Academician \\ Anatolii
Alekseevich Logunov} 
\end{center} 

%{}~

\newpage

%%%%%%%%%%%%%%%%%%%%%%%%%%

\pagestyle{plain}
\renewcommand{\thefootnote}{\arabic{footnote}}
\setcounter{footnote}{0}

\section{Introduction}

In several recent decades, quantum field theory, having incorporated
efficient mathematical methods, has become a theory satisfying the most
rigorous mathematical requirements~\cite{1}. After the formulation of
supersymmetric quantum field theories, the main attention of investigators
was attracted to numerous problems whose solution is interesting in both
mathematical physics and important physical applications.

In this paper, we consider an integrable $N{=}(1|1)$ supersymmetric
generalization of the two-dimensional bosonic Toda lattice hierarchy
(2DTL hierarchy)~\cite{2} proposed in~\cite{3},~\cite{4}. It is given by an
infinite system of evolution equations (flows) for an infinite set of bosonic
and fermionic lattice fields evolving in two bosonic and two fermionic
infinite ``towers" of times; as a subsystem, it involves an $N{=}(1|1)$
supersymmetric integrable generalization of the 2DTL equation, which is
called the $N{=}(1|1)$ 2DTL in what follows.

Two new infinite series of fermionic flows of the $N{=}(1|1)$ 2DTL hierarchy
were constructed in~\cite{5}--\cite{7}. This hierarchy was shown to actually
have a higher symmetry, namely, the $N=(2|2)$ supersymmetry. Together with
the previously known bosonic flows of the $N{=}(1|1)$ 2DTL hierarchy, these
flows are symmetries of the $N{=}(1|1)$ 2DTL equation. The continuum limit of
that equation with respect to the lattice constant~\cite{8} is a
three-dimensional nonlinear equation, called the continuum or dispersionless
$N{=}(1|1)$ 2DTL equation. The solution of the corresponding Cauchy problem
was also considered in~\cite{8}.

Although the $N{=}(1|1)$ 2DTL hierarchy and the dispersionless $N{=}(1|1)$
2DTL equation have been known for a relatively long time, the problem of
constructing the continuum (semiclassical) limit with respect to 
the lattice
constant (which plays the role of the Planck constant here) for all the
$N{=}(1|1)$ 2DTL hierarchy flows is still to be solved and is quite
complicated.\footnote{We note that similar problems for
differential--difference equations were considered in~\cite{9}--\cite{16}.}
Apart from the purely academic significance of this problem, its solution is
also interesting in relation to a number of important physical and
mathematical applications. In particular, these include the semiclassical
limit of the bosonic preimage of the $N{=}(1|1)$ 2DTL hierarchy, the
dispersionless 2DTL hierarchy~\cite{17}, which unifies the 2DTL hierarchy
flows arising in the leading semiclassical approximation constructed
in~\cite{18} (see also~\cite{19}). Possible applications of the
dispersionless 2DTL hierarchy are

1. construction of a number of self-dual vacuum metrics and
Einstein--Weyl metrics,

2. twistor theory,

3. two-dimensional conformal and topological field theory, and

4. two-dimensional string theory

(also see~\cite{20}--\cite{31} and the references therein).

In view of a deep relation between the 2DTL and $N{=}(1|1)$ 2DTL hierarchies,
it is natural to hypothesize that the dispersionless $N{=}(1|1)$
supersymmetric 2DTL hierarchy must also admit similar applications in
supersymmetric generalizations of the theories listed above. This motivates
our construction of the dispersionless $N{=}(1|1)$ 2DTL hierarchy here.

We emphasize that the existing general algorithm for constructing
semiclassical asymptotic expansions, which was applied to the 2DTL hierarchy
in~\cite{18}, encounters a number of difficulties, both formal and informal,
in its direct extension to the $N{=}(1|1)$ 2DTL hierarchy. It is well-known
that in the semiclassical limit, all the operators are mapped into their
symbols defined on the corresponding phase space and the (anti)commutators
involved in the standard Lax representations are replaced with the
corresponding Poisson brackets. But the Lax representation of the $N{=}(1|1)$
2DTL hierarchy proposed in~\cite{3} is not of the (anti)commutator form, and
it is therefore unclear how the above qualitative recipe for the transition
to the semiclassical regime can be applied literally.

The analysis is also complicated by another principally important,
qualitatively new feature of the $N{=}(1|1)$ 2DTL hierarchy compared with its
bosonic preimage, namely, the operators entering the Lax representation are
defined on a space with two $Z_2$-gradings but have only one diagonal
$Z_2$-grading. This is why the symbols of these operators in the general case
cannot be expected to commute, even in the semiclassical limit.

We also mention possible complications related to the fact that the fermionic
and bosonic fields of the $N{=}(1|1)$ 2DTL hierarchy can have noncoincident
semiclassical asymptotic behaviors, which must then be specified
self-consistently. A similar situation occurs for the fermionic and bosonic
times of this hierarchy, but these can be easily made consistent using
additional dimension arguments.

Very recently, a certain progress has been observed toward constructing the
semiclassical limit of the $N{=}(1|1)$ 2DTL hierarchy. In~\cite{7}, it was
found that the Lax representation for this hierarchy can be brought to the
commutator form (Eq.~(39) in~\cite{7}) by introducing a number of new
auxiliary fermionic constants. Although the commutator representation thus
obtained seems artificial at first glance, it actually involves a new
important generalized graded bracket operation, which can be defined in
sufficiently general terms to allow a broad spectrum of applications. We
introduce and use it in solving the problem addressed here.

It turns out that all the basic relations defining the $N{=}(1|1)$ 2DTL
hierarchy can be expressed in terms of precisely this bracket operation; this
does not require introducing any auxiliary objects like the fermionic
constants mentioned above. In the semiclassical limit, moreover, precisely
this bracket operation is replaced with the corresponding Poisson superbracket
on the phase superspace. This therefore produces the Lax representation of
the dispersionless $N{=}(1|1)$ 2DTL hierarchy.

The structure of this paper is as follows. In Sec.~2, we introduce the
generalized graded bracket operation on the space of graded operators with an
involution that generalizes the graded commutator for superalgebras; we
describe its properties and give the corresponding generalized Jacobi
identities. We then obtain the Lax representation of the $N{=}(1|1)$ 2DTL
hierarchy and all the basic defining relations in terms of the generalized
graded bracket. We give an explicit expression for the $N{=}(1|1)$ 2DTL
hierarchy flows and for bosonic symmetries of the $N{=}(1|1)$ 2DTL equation,
which are consequently used in Sec.~3 to obtain the respective dispersionless
analogues.

In Sec.~3, we also find the semiclassical limit of the $N{=}(1|1)$ 2DTL
hierarchy and postulate the corresponding asymptotic behavior of the
fermionic and bosonic fields parameterizing the Lax operators. Using these
data, we then evaluate the asymptotic behavior of all the composite operators
entering the Lax representation and the corresponding field evolution
equations. We next obtain regular leading terms in the semiclassical
expansion of these evolution equations, which are by definition the flows of
the dispersionless $N{=}(1|1)$ 2DTL hierarchy. This a posteriori demonstrates
self-consistency of the postulates underlying all our calculations. The next
step is to model the Poisson superbracket on the phase superspace obtained by
extending the phase space of the dispersionless 2DTL hierarchy by one
Grassmann coordinate. Replacing the Lax operators with their symbols and
replacing the generalized graded bracket with the above Poisson superbracket
in the Lax representation of the $N{=}(1|1)$ 2DTL hierarchy and in all its
defining relations, we show by a direct calculation that the operator
representation thus obtained correctly reproduces the flows of the
dispersionless $N{=}(1|1)$ 2DTL hierarchy that we constructed previously and
is therefore the sought Lax representation of the above hierarchy.

In Sec.~4, we briefly summarize the main results obtained in this paper.

\section{$N{=}(1|1)$ 2DTL hierarchy}

In this section, we introduce a new graded bracket operation and use it to
propose a new form of the Lax representation for the $N{=}(1|1)$ 2DTL
hierarchy.   

\subsection{Generalized graded brackets}

We consider the space of operators ${\bO}_k$ with the grading 
$d_{{\bO}_k}$  ($d_{{\bO}_k}\in {\bZ}$),  
\begin{eqnarray}
d_{{\bO}_1{\bO}_2}= d_{{\bO}_1}+d_{{\bO}_2}, \label{grad}
\end{eqnarray}
and the involution $*$, 
\begin{eqnarray}
{\bO}^{*(2)}_k = {\bO}_k, \label{inv}
\end{eqnarray}
where  ${{\bO}_k}^{*(m)}$  denotes the $m$-fold action of the involution 
$*$ on the operator ${\bO}_k$. On this space, we can define the 
generalized graded bracket operation $[\,\cdot\,,\,\cdot\,\}$, 
\begin{eqnarray}
\Bigl[ {\bO}_1, {\bO}_2 \Bigr\}:={\bO}_1{\bO}_2
 - (-1)^{d_{{\bO}_1}d_{{\bO}_2}}~{{\bO}_2}^{*(d_{{\bO}_1})}~
{{\bO}_1}^{*(d_{{\bO}_2})}
  \label{brack}
\end{eqnarray}
with the easily verified properties of

{\sl a.  symmetry}               
\begin{eqnarray}
\Bigl[ {\bO}_1, {\bO}_2 \Bigr\}=(-1)^{d_{{\bO}_1}d_{{\bO}_2}+1}
\Bigl[{{\bO}_2}^{*(d_{{\bO}_1})},{{\bO}_1}^{*(d_{{\bO}_2})}
\Bigr\}, \label{prop1}
\end{eqnarray}

{\sl b. derivation}    
\begin{eqnarray}
\Bigl[ {\bO}_1, {\bO}_2{\bO}_3\Bigr\}=
\Bigl[{\bO}_1,{\bO}_2\Bigr\}{\bO}_3 +
(-1)^{d_{{\bO}_1}d_{{\bO}_2}}{{\bO}_2^{*(d_{{\bO}_1})}}
\Bigl[{{\bO}_1}^{*(d_{{\bO}_2})},{\bO}_3 \Bigr\}, \label{prop2}
\end{eqnarray}

{\sl c. Jacobi identity} 
\begin{eqnarray}
&&(-1)^{d_{{\bO}_1}d_{{\bO}_3}}\Bigl[\Bigl[
{{\bO}_1}^{*(d_{{\bO}_3)}}, {\bO}_2
\Bigr\},{{\bO}_3}^{*(d_{{\bO}_1)}}\Bigr\} \nonumber\\
&+&(-1)^{d_{{\bO}_2}d_{{\bO}_1}} \Bigl[\Bigl[
{{\bO}_2}^{*(d_{{\bO}_1)}},
{\bO}_3\Bigr\},{{\bO}_1}^{*(d_{{\bO}_2)}}\Bigr\} \nonumber\\&+&
(-1)^{d_{{\bO}_3}d_{{\bO}_2}} \Bigl[\Bigl[
{{\bO}_3}^{*(d_{{\bO}_2)}},
{\bO}_1\Bigr\},{{\bO}_2}^{*(d_{{\bO}_3)}}\Bigr\}=0. \label{jacobi}
\end{eqnarray}
Relations (\ref{prop1}--\ref{jacobi}) generalize the corresponding 
properties of the graded commutator in Lie superalgebras. We emphasize 
that in the particular case where the involution $*$ acts as the
identity transformation, bracket \p{brack} reproduces the graded Lie
superalgebra commutator. In the general case of the involution action, 
this bracket is a nontrivial generalization of that commutator.            

\subsection{Lax representation and flows}

We begin this section by detailing the space of operators, their grading, and
involution which are relevant in context of the $N{=}(1|1)$ 2DTL hierarchy. 
These operators can be represented in the general form 
\begin{eqnarray}
{\bO}_m=\sum^{\infty}_{k=-\infty}
f^{(m)}_{k,j}e^{(k-m){\partial}}, \quad m\in {\bZ}, \label{optoda}
\end{eqnarray}
parameterized by the functions $f^{(m)}_{2k,j}$ ($f^{(m)}_{2k+1,j}$) 
that are $Z_2$-graded bosonic (fermionic) lattice fields ($j \in {\bZ}$),
\begin{eqnarray}
d^{~'}_{f^{(m)}_{k,j}}=|k| \quad \mbox{mod} ~~ 2; \label{grad1}
\end{eqnarray}
the operator $e^{l{\partial}}$ ($l \in {\bZ}$) acting on these fields as 
the discrete lattice shift                                                               
\begin{eqnarray}
e^{l{\partial}} f^{(m)}_{k,j} \equiv f^{(m)}_{k,j+l}
e^{l{\partial}} \label{rule1}
\end{eqnarray}
has another $Z_2$-grading given by 
\begin{eqnarray}
d_{e^{l{\partial}}}=|l| \quad \mbox{mod}  ~~ 2. \label{grad2}
\end{eqnarray}
Operators \p{optoda} allow specifying only one diagonal 
$Z_2$-grading 
\begin{eqnarray}
d_{{\bO}_m}=d^{~'}_{f^{(m)}_{k,j}}+d_{e^{(k-m){\partial}}} =|m|
\quad \mbox{mod}  ~~ 2
 \label{gradop}
\end{eqnarray}
and the involution 
\begin{eqnarray}
{\bO}^{*}_m=\sum^{\infty}_{k=-\infty} (-1)^k
f^{(m)}_{k,j}e^{(k-m){\partial}}. \label{invtoda}
\end{eqnarray}
In what follows, we also need the projections ${({\bO}_m)}_{\pm}$
of the operators ${\bO}_m$ \p{optoda} defined as 
\begin{eqnarray}
({\bO}_m)_{+}=\sum^{\infty}_{k=m}
f^{(m)}_{k,j}e^{(k-m){\partial}}, \quad
{({\bO}_m)}_{-}=\sum^{m-1}_{k=-\infty}
f^{(m)}_{k,j}e^{(k-m){\partial}}.
 \label{optoda1}
\end{eqnarray}

The Lax operators $L^{\pm}$ of the $N{=}(1|1)$ 2DTL hierarchy belong to 
the space of operators \p{optoda} \cite{3,7}
\begin{eqnarray}
L^{+}=\sum^{\infty}_{k=0} u_{k,j}e^{(1-k){\partial}},\quad
u_{0,j}= 1, \quad L^{-}=\sum^{\infty}_{k=0}
v_{k,j}e^{(k-1){\partial}}, \quad v_{0,j}\neq 0 \label{laxs1}
\end{eqnarray}
and have the grading $d_{L^{\pm}}=1$.

We now have all the ingredients necessary to express the Lax 
representation of the $N{=}(1|1)$ 2DTL hierarchy in terms of bracket 
operation \p{brack}, thereby bringing it to a very simple form, 
\begin{eqnarray}
D^{\pm}_n L^{{\alpha}}&=& \mp {\alpha} (-1)^n
\Bigr[{(((L^{\pm})^{n}_{*})}_{-{\alpha} })^{*},
L^{{\alpha}}\Bigl\}, \quad {\alpha} =+,-, \quad n \in {\bN},
\label{laxreprs1}
\end{eqnarray}
where $D^{\pm}_{2n}$ ($D^{\pm}_{2n+1}$) are bosonic (fermionic) evolution
derivatives.                                         

For the composite operators $(L^{\pm})^{n}_{*}$ entering this 
representation, we can also obtain very simple expressions in terms of the 
Lax operators and bracket operation \p{brack},
\begin{eqnarray}
(L^{{\alpha}})^{2n}_{*}:= \Bigl(~\frac{1}{2}~\Bigr[
(L^{{\alpha}})^{*}, (L^{{\alpha}})\Bigl\}~\Bigr)^n, \quad
(L^{{\alpha}})^{2m+1}_{*}:=L^{{\alpha}}~(L^{{\alpha}})^{2n}_{*}.
\label{rule2}
\end{eqnarray}
Similarly to the Lax operators $L^{\pm}$, the operators 
$(L^{\pm})^{n}_{*}$ belong to the space of operators \p{optoda} 
and can be represented as 
\begin{eqnarray}
(L^{+})^{m}_{*}:=\sum^{\infty}_{k=0}
u^{(m)}_{k,j}e^{(m-k){\partial}}, \quad u^{(m)}_{0,j}=1, \quad
(L^{-})^{m}_{*}:=\sum^{\infty}_{k=0}
v^{(m)}_{k,j}e^{(k-m){\partial}}, \label{laxs2}
\end{eqnarray}
where  $u^{(m)}_{k,j}$ and $v^{(m)}_{k,j}$ (with $u^{(1)}_{k,j}\equiv
u_{k,j},~v^{(1)}_{k,j}\equiv v_{k,j}$) are functionals of the original 
fields $\{u_{k,j},~v_{k,j}\}$. It must be noted here that in Lax
representation \p{laxreprs1}, the  $Z_2$-grading of the operatorа
$(L^{\pm})^{n}_{*}$, which has the form $d_{(L^{\pm})^{2n}_{*}}=0$ and 
$d_{(L^{\pm})^{2n+1}_{*}}=1$, agrees with another $Z_2$-grading 
$d^{~'}_{D^{\pm}_{2n}}=0$ and $d^{~'}_{D^{\pm}_{2n+1}}=1$ 
that corresponds to the statistics of the evolution derivatives 
$D^{\pm}_{n}$.

Using bracket properties (\ref{prop1}--\ref{jacobi}) and relations
(\ref{rule2}) as definitions of $(L^{\pm})^{n}_{*}$, we can easily obtain 
the useful identities   
\begin{eqnarray}
&&\Bigr[ (L^{{\alpha}})^{2n}_{*},
(L^{{\alpha}})^{2m}_{*}\Bigl\}=0, \nonumber\\
&&\Bigr[ ((L^{{\alpha}})^{2n}_{*})^{*},
(L^{{\alpha}})^{2m+1}_{*}\Bigl\}=0, \quad \Bigr[
(L^{{\alpha}})^{2n+1}_{*},
(L^{{\alpha}})^{2m}_{*}\Bigl\}=0, \nonumber\\
&&\Bigr[ ((L^{{\alpha}})^{2n+1}_{*})^{*},
(L^{{\alpha}})^{2m+1}_{*}\Bigl\}=2(L^{{\alpha}})^{2(n+m+1)}_{*}.
\label{ident}
\end{eqnarray}
Next, using Eqs. (\ref{prop1}--\ref{jacobi}) 
and (\ref{laxreprs1}--\ref{rule2}), we can derive equations of motion for 
the composite operators $(L^{\pm})^{n}_{*}$,
\begin{eqnarray}
D^{\pm}_n (L^{{\alpha}})^{m}_{*}&=& \mp {\alpha} (-1)^{nm} \Bigr[
{(((L^{\pm})^{n}_{*})}_{-{\alpha} })^{*(m)},
(L^{{\alpha}})^{m}_{*}\Bigl\}, \label{laxreprs2}
\end{eqnarray}
and the evolution equations for the functionals 
$\{u^{(m)}_{k,j},~v^{(m)}_{k,j}\}$ in \p{laxs2} implied by Eqs. 
\p{laxreprs2} found above, 
\begin{eqnarray}
D^{+}_n u^{(2m)}_{k,j}&=&\sum^{n}_{p=0}
(u^{(n)}_{p,j}u^{(2m)}_{k-p+n,j-p+n} \nonumber\\
&-&(-1)^{(p+n)(k-p+n)}u^{(n)}_{p,j-k+p-n+2m}u^{(2m)}_{k-p+n,j}),
\nonumber\\ D^{+}_{2n} u^{(2m+1)}_{k,j}&=&\sum^{2n}_{p=0}
((-1)^{p}u^{(2n)}_{p,j}u^{(2m+1)}_{k-p+2n,j-p+2n} \nonumber\\
&-&(-1)^{p(k-p)}u^{(2n)}_{p,j-k+p-2n+2m+1}u^{(2m+1)}_{k-p+2n,j}),
\nonumber\\ D^{+}_{2n+1} u^{(2m+1)}_{k,j}&=&\sum^{k}_{p=1}
((-1)^{p+1}u^{(2n+1)}_{p+2n+1,j}u^{(2m+1)}_{k-p,j-p} \nonumber\\
&+&(-1)^{p(k-p)}u^{(2n+1)}_{p+2n+1,j-k+p+2m+1}u^{(2m+1)}_{k-p,j}),
\label{equationss1}
\end{eqnarray}
\begin{eqnarray}
D^{-}_n u^{(m)}_{k,j}&=&\sum^{n-1}_{p=0}
((-1)^{(p+n)m}v^{(n)}_{p,j}u^{(m)}_{k+p-n,j+p-n} \nonumber\\
&-&(-1)^{(p+n)(k+p-n)}v^{(n)}_{p,j-k-p+n+m}u^{(m)}_{k+p-n,j}),
\label{equationss2}
\end{eqnarray}
\begin{eqnarray}
D^{+}_n v^{(m)}_{k,j}&=&\sum^{n}_{p=0}
((-1)^{(p+n)m}u^{(n)}_{p,j}v^{(m)}_{k+p-n,j-p+n} \nonumber\\
&-&(-1)^{(p+n)(k+p-n)}u^{(n)}_{p,j+k+p-n-m}v^{(m)}_{k+p-n,j}),
\label{equationss3}
\end{eqnarray}
\begin{eqnarray}
D^{-}_n v^{(2m)}_{k,j}&=&\sum^{n-1}_{p=0}
(v^{(n)}_{p,j}v^{(2m)}_{k-p+n,j+p-n} \nonumber\\
&-&(-1)^{(p+n)(k-p+n)}v^{(n)}_{p,j+k-p+n-2m}v^{(2m)}_{k-p+n,j}),
\nonumber\\ D^{-}_{2n} v^{(2m+1)}_{k,j}&=&\sum^{2n-1}_{p=0}
((-1)^{p}v^{(2n)}_{p,j}v^{(2m+1)}_{k-p+2n,j+p-2n} \nonumber\\
&-&(-1)^{p(k-p)}v^{(2n)}_{p,j+k-p+2n-2m-1}v^{(2m+1)}_{k-p+2n,j}),
\nonumber\\ D^{-}_{2n+1} v^{(2m+1)}_{k,j}&=&\sum^{k}_{p=0}
((-1)^{p+1}v^{(2n+1)}_{p+2n+1,j}v^{(2m+1)}_{k-p,j+p} \nonumber\\
&+&(-1)^{p(k-p)}v^{(2n+1)}_{p+2n+1,j+k-p-2m-1}v^{(2m+1)}_{k-p,j})
\label{equationss4b}
\end{eqnarray}
(in the right-hand side of these equations, all the fields  
$\{u^{(m)}_{k,j},~v^{(m)}_{k,j}\}$ with $k < 0$  must be set equal to 
zero).

Lax representation \p{laxreprs1} generates a non-Abelian algebra of flows 
of the $N{=}(1|1)$ 2DTL hierarchy, 
\begin{eqnarray}
[D^{+}_{n}~,~D^{-}_{l}\}=[D^{\pm}_{n}~,~D^{\pm}_{2l}]=0,\quad
\{D^{\pm}_{2n+1}~,~D^{\pm}_{2l+1}\}=2D^{\pm}_{2(n+l+1)},
\label{algebras1}
\end{eqnarray}
which can be realized as    
\begin{eqnarray}
D^{\pm}_{2n} ={\partial}^{\pm}_{2n}, \quad D^{\pm}_{2n+1}
={\partial}^{\pm}_{2n+1}+
\sum^{\infty}_{l=1}t^{\pm}_{2l-1}{\partial}^{\pm}_{2(k+l)},\quad
\partial^{\pm}_n:={\frac{\partial}{\partial t^{\pm}_n}},
\label{covder}
\end{eqnarray}
where $t^{\pm}_{2n}$ ($t^{\pm}_{2n+1}$) 
are bosonic (fermionic) evolution times.
The dispersionless $N{=}(1|1)$ 2DTL hierarchy flows are based on the
continuum limit of flows (\ref{equationss1}--\ref{equationss4b}) 
constructed in Sec.~3.1. 

\subsection{Bosonic symmetries of the $N{=}(1|1)$ 2DTL equation}

The supersymmetric $N{=}(1|1)$ 2DTL equation 
\begin{eqnarray}
D^{+}_1D^{-}_1 \ln v_{0,j}= v_{0,j+1} - v_{0,j-1} \label{todabs}
\end{eqnarray}
belongs to system of equations (\ref{equationss1}--\ref{equationss4b}). 
It can be obtained from Eq. \p{equationss2} with $\{n=m=k=1\}$,
\begin{eqnarray}
D^{-}_1 u_{1,j} = - v_{0,j} - v_{0,j+1}, \label{todabs1}
\end{eqnarray}
and from Eq. \p{equationss3} with $\{n=m=1,~ k=0\}$,
\begin{eqnarray}
D^{+}_1 v_{0,j}= v_{0,j}(u_{1,j} - u_{1,j-1}), \label{todabs2}
\end{eqnarray}
after eliminating the field $u_{1,j}$. Its bosonic symmetries 
$D^{\pm}_{2n} v_{0,j}$, with
\begin{eqnarray}
\{D^{+}_1,D^{\pm}_{2n}\}= \{D^{-}_1,D^{\pm}_{2n}\}=0,
\label{algebra2new}
\end{eqnarray}
were described in~\cite{5}--\cite{7} in terms of the iteration procedure  
\begin{eqnarray}
D^{\pm}_{2n}
v_{0,j}&=&v_{0,j}(u^{(2n)\pm}_{2n,j}-u^{(2n)\pm}_{2n,j-1}), \quad
u^{(2n)\pm}_{0,j}=1, \nonumber\\ \pm D^{\mp}_1
u^{(2n)\pm}_{k,j}&=&v_{0,j}u^{(2n)\pm}_{k-1,j-1} +
(-1)^{k}v_{0,j-k+2n+1}u^{(2n)\pm}_{k-1,j}, \label{flowss+-bos}
\end{eqnarray}
where the functions $u^{(n)\pm}_{k,j}$ are related to the original 
functionals $\{u^{(n)}_{k,j}, v^{(n)}_{k,j}\}$ as
\begin{eqnarray}
u^{(n)+}_{k,j}= u^{(n)}_{k,j}, \quad u^{(n)-}_{k,j}=
\frac{v^{(n)}_{k,-j-1}}{{\sum}^{n-k}_{m=1} v_{0,k+m-n-j-1}}
\label{relation}
\end{eqnarray}
(for detail, see \cite{7}).

The continuum limit of $N{=}(1|1)$ 2DTL equation \p{todabs} and of its
symmetries $D^{\pm}_{2n} v_{0,j}$ \p{flowss+-bos} is derived in Sec.~3.2.

\section{The dispersionless $N{=}(1|1)$ 2DTL hierarchy}

In this section, we construct the continuum (semiclassical) limit of the
$N{=}(1|1)$ 2DTL hierarchy with respect to the lattice constant, which 
gives the dispersionless $N{=}(1|1)$ 2DTL hierarchy, and construct the
corresponding Lax representation.             

\subsection{Semiclassical limit}

Flows (\ref{equationss1}--\ref{equationss4b}) of the $N{=}(1|1)$ 2DTL 
hierarchy do not involve an explicit dependence on dimensional constants, 
and the lattice with the dimensionless coordinate $j$ ($j\in {\bZ})$ 
has the unit spacing constant. To study the continuum limit, we explicitly 
introduce the lattice spacing length. This parameter is denoted by $\hbar$ 
because it plays the role of the Planck constant in what follows. Instead 
of $j$, there then arises the combination                            
\begin{eqnarray}
{\hbar} j \equiv s, \label{coord1}
\end{eqnarray}
and all the lattice fields acquire a dependence on the parameter $\hbar$. 
The continuum (semiclassical) limit can then be defined as   
\begin{eqnarray}
{\hbar}\rightarrow 0, \quad s = {\lim}_{{\hbar}\rightarrow 0, ~j
>> 1}({\hbar} j), \label{displimit1}
\end{eqnarray}
with $s$ playing the role of the continuum ``lattice" coordinate.

For the flows in Eqs. (\ref{equationss1}--\ref{equationss4b})
to be nontrivial and regular in the limit in Eq. \p{displimit1},
we must additionally perform several scaling transformations of the 
dependent and independent variables in the system. We postulate the rules 
for transition to the new evolution times and fields of the hierarchy 
given by                                            
\begin{eqnarray}
t^{\pm}_{2n+1} \rightarrow \frac{1}{\sqrt{\hbar}} t^{\pm}_{2n+1},
~~ t^{\pm}_{2n} \rightarrow \frac{1}{{\hbar}} t^{\pm}_{2n}
~\Leftrightarrow ~ D^{\pm}_{2n+1}  \rightarrow \sqrt{\hbar}
D^{\pm}_{2n+1}, ~~ D^{\pm}_{2n}  \rightarrow {\hbar}
D^{\pm}_{2n},~~~~~~ \label{displimit2}
\end{eqnarray}
\begin{eqnarray}
&&u_{2k,j} \quad ~\rightarrow ~~~~u_{2k}({\hbar} j), \quad
\quad ~~~v_{2k,j} ~\quad \rightarrow ~~~~v_{2k}({\hbar} j),\nonumber\\
&&u_{2k+1,j}~\rightarrow ~ \frac{1}{\sqrt{\hbar}}~u_{2k+1}({\hbar}
j), \quad ~~v_{2k+1,j}~~\rightarrow ~
\frac{1}{\sqrt{\hbar}}~v_{2k+1}({\hbar} j), \label{displimit3}
\end{eqnarray}
and assume they are nonsingular at $\hbar=0$. 

We can now establish two important properties of semiclassical limit 
(\ref{displimit1}--\ref{displimit3}), which have a key importance in 
what follows and are verified by direct calculations.   

The first property is that the new composite fields defined in accordance
with the rules    
\begin{eqnarray}
&&u^{(m)}_{2k,j} \quad ~\rightarrow ~~~~u^{(m)}_{2k}({\hbar} j),
\quad \quad ~v^{(m)}_{2k,j} \quad \rightarrow
~~~~~v^{(m)}_{2k}({\hbar} j), \nonumber\\
&&u^{(2m+1)}_{2k+1,j}\rightarrow ~\frac{1}{\sqrt{\hbar}}~
u^{(2m+1)}_{2k+1}({\hbar}j), \quad v^{(2m+1)}_{2k+1,j}\rightarrow
~ \frac{1}{\sqrt{\hbar}}~v^{(2m+1)}_{2k+1}({\hbar} j),\nonumber\\
&&u^{(2m)}_{2k+1,j}~\rightarrow ~\sqrt{\hbar}~
u^{(2m)}_{2k+1}({\hbar} j), \quad \quad
v^{(2m)}_{2k+1,j}~\rightarrow ~\sqrt{\hbar}~
v^{(2m)}_{2k+1}({\hbar}j), \label{displimit4}
\end{eqnarray}
are regular in the semiclassical limit.   

From rules (\ref{displimit1}--\ref{displimit4}) and the obvious 
identities  
\begin{eqnarray}
(L^{{\alpha}})^{2(m+1)}_{*}:= (L^{{\alpha}})^{2}_{*}
(L^{{\alpha}})^{2m}_{*}, \quad (L^{{\alpha}})^{(2m+1)}_{*}:=
L^{{\alpha}} (L^{{\alpha}})^{2m}_{*}, \label{ident1}
\end{eqnarray}
which follow from Eqs. \p{rule2}, we can, for example, obtain the important
recursive relations for the leading terms of the functionals  
$u^{(m)}_{k}\equiv u^{(m)}_{k}(s)$,
\begin{eqnarray}
&& u^{(2(l+1))}_{2k}
=\sum^{k}_{n=0}u^{(2)}_{2n}u^{(2l)}_{2(k-n)},\quad
u^{(2(l+1))}_{2k+1}=\sum^{2k+1}_{n=0}u^{(2)}_{n}u^{(2l)}_{2k-n+1},
\nonumber\\
&& u^{(2l+1)}_{2k} ~=\sum^{2k}_{n=0}u_{n}u^{(2l)}_{2k-n},\quad
\quad ~~u^{(2l+1)}_{2k+1}=\sum^{k}_{n=0}u_{2n+1}u^{(2l)}_{2(k-n)},
\label{rec1}
\end{eqnarray}
\begin{eqnarray}
&& u^{(2)}_{2k} =\sum^{k}_{n=0}u_{2n}u_{2(k-n)} +2\sum^{k-1}_{n=0}
(k-n-1)u_{2(k-n)-1}{\partial}_su_{2n+1}, \nonumber\\
&& u^{(2)}_{2k+1} =\sum^{k}_{n=0}\Big[(1-2n)u_{2n}{\partial}_s
u_{2(k-n)+1} +2(k-n)u_{2(k-n)+1} {\partial}_su_{2n}\Big],
\label{rec2}
\end{eqnarray}
where ${\partial}_s:= \frac{{\partial}}{{\partial s}}$.

The second property is that in semiclassical limit 
(\ref{displimit1}--\ref{displimit4}) flows 
(\ref{equationss1}--\ref{equationss4b}) of the $N{=}(1|1)$ 2DTL hierarchy 
are nontrivial and regular. Explicit expressions for their leading terms 
are as follows.                                        

For $\{m=2l, ~k=2r\}$ or $\{m=2l+1, ~k=2r+1\}$, we have 
\begin{eqnarray}
%при \quad \{m=2l, ~&k&=2r\}, \quad или \quad \{m=2l+1, ~k=2r+1\} 
%\nonumber\\
D^{-}_{2n+1} u^{(m)}_{k}&=&\sum^{n-1}_{p=0}
\Bigl[2(p-n)v^{(2n+1)}_{2p+1}{\partial}_s u^{(m)}_{k+2(p-n)}\nonumber\\
&+&(k-m+2(p-n))({\partial}_sv^{(2n+1)}_{2p+1})u^{(m)}_{k+2(p-n)}\Bigr]
\nonumber\\
&+&2(-1)^k\sum^{n}_{p=0}v^{(2n+1)}_{2p}u^{(m)}_{k+2(p-n)-1},
\nonumber\\
D^{-}_{2n} u^{(m)}_{k}&=&\sum^{n-1}_{p=0}
\Bigl[2(p-n)v^{(2n)}_{2p}{\partial}_s u^{(m)}_{k+2(p-n)}\nonumber\\
&+&(k-m+2(p-n))({\partial}_sv^{(2n)}_{2p})u^{(m)}_{k+2(p-n)}
\nonumber\\
&+&2(-1)^kv^{(2n)}_{2p+1}u^{(m)}_{k+2(p-n)+1}\Bigr], \nonumber\\
D^{+}_{2n} u^{(m)}_{k}&=&\sum^{n}_{p=0}
\Bigl[2(n-p)u^{(2n)}_{2p}{\partial}_s u^{(m)}_{k+2(n-p)}\nonumber\\
&+&(k-m+2(n-p))({\partial}_su^{(2n)}_{2p})u^{(m)}_{k+2(n-p)}\Bigr]
\nonumber\\&+& 2(-1)^k \sum^{n-1}_{p=0}
u^{(2n)}_{2p+1}u^{(m)}_{k+2(n-p)-1}, \nonumber\\
D^{+}_{2n+1} v^{(m)}_{k}&=&\sum^{n}_{p=0}
\Bigl[2(n-p)u^{(2n+1)}_{2p+1}{\partial}_s v^{(m)}_{k+2(p-n)}\nonumber\\
&-&(k-m+2(p-n))({\partial}_su^{(2n+1)}_{2p+1})v^{(m)}_{k+2(p-n)} \nonumber\\
&+& 2(-1)^k u^{(2n+1)}_{2p}v^{(m)}_{k-1+2(p-n)}\Bigr], \nonumber\\
D^{+}_{2n} v^{(m)}_{k}&=&\sum^{n}_{p=0}
\Bigl[2(n-p)u^{(2n)}_{2p}{\partial}_s v^{(m)}_{k+2(p-n)}\nonumber\\
&-&(k-m+2(p-n))({\partial}_su^{(2n)}_{2p})v^{(m)}_{k+2(p-n)}\Bigr]
\nonumber\\&+& 2(-1)^k \sum^{n-1}_{p=0}
u^{(2n)}_{2p+1}v^{(m)}_{k+1+2(p-n)}, \nonumber\\
D^{-}_{2n} v^{(m)}_{k}&=&\sum^{n-1}_{p=0}
\Bigl[2(p-n)v^{(2n)}_{2p}{\partial}_s v^{(m)}_{k+2(n-p)}\nonumber\\
&-&(k-m+2(n-p))({\partial}_sv^{(2n)}_{2p})v^{(m)}_{k+2(n-p)}\nonumber\\
&+&2(-1)^kv^{(2n)}_{2p+1}v^{(m)}_{k+2(n-p)-1}\Bigr].
\label{equ1+}
\end{eqnarray}
For $\{m=2l, ~k =2r+1\}$ or $\{m=2l+1, ~k=2r\}$, we have 
\begin{eqnarray}
%при \quad \{m=2l, ~&k&=2r+1\}, \quad или \quad \{m=2l+1, ~k=2r\} 
%\nonumber\\
D^{-}_n u^{(m)}_{k}&=&\sum^{n-1}_{p=0}
(-1)^{m(p+n)}\Bigl[(p-n)v^{(n)}_{p}{\partial}_s u^{(m)}_{k+p-n}\nonumber\\
&+&(k+p-n-m)({\partial}_sv^{(n)}_{p})u^{(m)}_{k+p-n}\Bigr],
\nonumber\\ D^{+}_{2n} u^{(m)}_{k}&=&\sum^{2n}_{p=0}
(-1)^{mp}\Bigl[(2n-p)u^{(2n)}_{p}{\partial}_s u^{(m)}_{k-p+2n}\nonumber\\
&+&(k-p+2n-m)({\partial}_su^{(2n)}_{p})u^{(m)}_{k-p+2n}\Bigr],
\nonumber\\ D^{+}_{n} v^{(m)}_{k}&=&\sum^{n}_{p=0}
(-1)^{m(p+n)}\Bigl[(n-p)u^{(n)}_{p}{\partial}_s v^{(m)}_{k+p-n}\nonumber\\
&-&(k+p-n-m)({\partial}_su^{(n)}_{p})v^{(m)}_{k+p-n}\Bigr],
\nonumber\\
D^{-}_{2n} v^{(m)}_{k}&=&\sum^{2n-1}_{p=0}
(-1)^{mp}\Bigl[(p-2n)v^{(2n)}_{p}{\partial}_s v^{(m)}_{k-p+2n}\nonumber\\
&-&(k-p+2n-m)({\partial}_sv^{(2n)}_{p})v^{(m)}_{k-p+2n}\Bigr],
\label{equ2+}
\end{eqnarray}
and also   
\begin{eqnarray}
D^{+}_{2n+1} u^{(2m)}_{2k}&=&2\sum^{n}_{p=0}
\Bigl[(n-p)u^{(2n+1)}_{2p+1}{\partial}_s u^{(2m)}_{2(k-p+n)}\nonumber\\
&+&(k-p+n-m)({\partial}_su^{(2n+1)}_{2p+1})u^{(2m)}_{2(k-p+n)}
\nonumber\\ &+& u^{(2n+1)}_{2p}u^{(2m)}_{2(k-p+n)+1}\Bigr],
\nonumber\\ D^{+}_{2n+1} u^{(2m+1)}_{2k+1}&=&2\sum^{k}_{p=1}
\Bigl[pu^{(2n+1)}_{2(p+n)+1}{\partial}_s u^{(2m+1)}_{2(k-p)+1}\nonumber\\
&+&(m+p-k)({\partial}_su^{(2n+1)}_{2(p+n)+1})u^{(2m+1)}_{2(k-p)+1}\Bigr]
\nonumber\\ &+& 2\sum^{k}_{p=0}
u^{(2n+1)}_{2(p+n+1)}u^{(2m+1)}_{2(k-p)},\nonumber\\
D^{+}_{2n+1} u^{(2m+1)}_{2k}&=&\sum^{2k}_{p=1}
(-1)^p\Bigl[pu^{(2n+1)}_{p+2n+1}{\partial}_s u^{(2m+1)}_{2k-p}\nonumber\\
&+&(2(m-k)+p+1)({\partial}_su^{(2n+1)}_{p+2n+1})u^{(2m+1)}_{2k-p}\Bigr],
\nonumber\\ D^{+}_{2n+1} u^{(2m)}_{2k+1}&=&\sum^{2n+1}_{p=0}
\Bigl[(2n-p+1)u^{(2n+1)}_{p}{\partial}_s u^{(2m)}_{2(k+n+1)-p}\nonumber\\
&+&(2(k+n-m+1)-p)({\partial}_su^{(2n+1)}_{p})u^{(2m)}_{2(k+n+1)-p}\Bigr],
\nonumber\\D^{-}_{2n+1} v^{(2m)}_{2k}&=&2\sum^{n-1}_{p=0}
\Bigl[(p-n)v^{(2n+1)}_{2p+1}{\partial}_s v^{(2m)}_{2(k-p+n)}\nonumber\\
&-&(k-p+n-m)({\partial}_sv^{(2n+1)}_{2p+1})v^{(2m)}_{2(k-p+n)}\Bigr]
\nonumber\\ &+&2\sum^{n}_{p=0}
v^{(2n+1)}_{2p}v^{(2m)}_{2(k-p+n)+1}, \nonumber\\D^{-}_{2n+1}
v^{(2m+1)}_{2k+1}&=&2\sum^{k}_{p=0}
\Bigl[-pv^{(2n+1)}_{2(p+n)+1}{\partial}_s v^{(2m+1)}_{2(k-p)+1}\nonumber\\
&+&(k-p-m){\partial}_sv^{(2n+1)}_{2(p+n)+1})v^{(2m+1)}_{2(k-p)+1}
\nonumber\\ &+& v^{(2n+1)}_{2(p+n+1)}v^{(2m+1)}_{2(k-p)}\Bigr],
\nonumber\\D^{-}_{2n+1} v^{(2m+1)}_{2k}&=&\sum^{k}_{p=0}
(-1)^p\Bigl[-pv^{(2n+1)}_{p+2n+1}{\partial}_s v^{(2m+1)}_{2k-p}\nonumber\\
&-&(2(m-k)+p+1)({\partial}_sv^{(2n+1)}_{p+2n+1})v^{(2m+1)}_{2k-p}\Bigr],
\nonumber\\D^{-}_{2n+1} v^{(2m)}_{2k+1}&=&\sum^{2n}_{p=0}
\Bigl[-(2n-p+1)v^{(2n+1)}_{p}{\partial}_s v^{(2m)}_{2(k+n+1)-p}\nonumber\\
&-&(2(k+n-m+1)-p)({\partial}_sv^{(2n+1)}_{p})v^{(2m)}_{2(k+n+1)-p}\Bigr],
\label{eqvv6-}
\end{eqnarray}
where all the fields $\{u^{(m)}_{k},~v^{(m)}_{k}\}$ with  $k < 0$
must be set equal to zero in the right-hand sides. Flows
(\ref{equ1+}--\ref{eqvv6-}) thus derived are said to constitute the 
dispersionless $N{=}(1|1)$ 2DTL hierarchy. The corresponding Lax 
representation is constructed in Sec.~3.3.            

\subsection{The dispersionless $N{=}(1|1)$ 2DTL equation and its bosonic
symmetries}

The dispersionless limits of $N{=}(1|1)$ 2DTL equation \p{todabs} and of 
its bosonic symmetries (\ref{flowss+-bos}) can be easily obtained using 
Eqs. (\ref{relation}--\ref{displimit4}). We respectively obtain 
\begin{eqnarray}
D^{+}_1D^{-}_1 \ln v_{0}= 2 {\partial}_s v_{0}
\label{disptoda}
\end{eqnarray}
and 
\begin{eqnarray}
D^{\pm}_{2n} v_{0}&=&v_{0}{\partial}_s
u^{(2n)\pm}_{2n}, \quad u^{(2n)\pm}_{0,j}=1, \nonumber\\
\mp D^{\mp}_1 u^{(2n)\pm}_{2k+1}&=&v_{0}{\partial}_s
u^{(2n)\pm}_{2k} + 2(n-k)({\partial}_sv_{0})u^{(2n)\pm}_{2k},
\nonumber\\ \pm D^{\mp}_1
u^{(2n)\pm}_{2k}&=&2v_{0}u^{(2n)\pm}_{2k-1}.
\label{disp-flowss+-bos}
\end{eqnarray}

Eliminating $u^{(2n)\pm}_{2k+1}$ from system of equations 
\p{disp-flowss+-bos}, we finally obtain a recursive system of equations 
for the generation of bosonic symmetries of dispersionless $N{=}(1|1)$ 
2DTL equation \p{disptoda}
\begin{eqnarray}
D^{\pm}_{2n} v_{0}&=&v_{0}{\partial}_s
u^{(2n)\pm}_{2n}, \quad u^{(2n)\pm}_{0,j}=1, \nonumber\\
-D^{\mp}_1 u^{(2n)\pm}_{2k}&=&2v_{0} {(D^{\mp}_1)}^{-1} \Bigl[
v_{0}{\partial}_s u^{(2n)\pm}_{2(k-1)} +
2(n-k+1)({\partial}_sv_{0})u^{(2n)\pm}_{2(k-1)}\Bigr].
\label{disp-flowss+-bos1}
\end{eqnarray}
We note that symmetries of the bosonic dispersionless 2DTL equation were
found in~\cite{32} (also see~\cite{26}) by solving the corresponding
sufficiently complicated symmetry equation.                           

\subsection{Lax representation}

In Sec.~3.1, we constructed flows (\ref{equ1+}--\ref{eqvv6-}) of the
dispersionless $N{=}(1|1)$ 2DTL hierarchy. We now find the corresponding 
Lax representation. Because Lax representation \p{laxreprs1} 
of the $N{=}(1|1)$ 2DTL hierarchy can be expressed in terms of generalized 
graded bracket \p{brack}, it is natural to expect that to derive its 
semiclassical limit, the corresponding dispersionless $N{=}(1|1)$ 2DTL 
hierarchy, it would suffice to replace all occurrences of this bracket 
with a certain Poisson superbracket and the operators with their symbols 
defined on the corresponding phase space.

Our immediate problem is to model the phase space and the Poisson    
superbracket starting from certain properties with which they must be
endowed. Recalling that the Lax operators of the $N{=}(1|1)$ 2DTL 
hierarchy, being defined on the space of operators graded by two different
$Z_2$-gradings (\ref{grad1}) and (\ref{grad2}), have only one diagonal
$Z_2$-grading \p{gradop}, we can assume that the phase space inherits these
properties. With the $Z_2$-grading $d_{e^{{\partial}}}=1, 
~d_{e^{2{\partial}}}=0$ of the lattice shift operator $e^{l{\partial}}$ 
taken into account, (\ref{grad2}), we assume that its counterpart on the 
phase space is given by two coordinates, the Grassmann coordinate $\pi$ 
(with $\pi^2=0$) and the bosonic coordinate $p$, via                                                                      
\begin{eqnarray}
e^{{\partial}}~\rightarrow ~ \frac{1}{{\sqrt\hbar}}~\pi, \quad
 e^{2{\partial}}~\rightarrow  ~p . \label{displax1}
\end{eqnarray}
Although the coordinate $\pi$ is Grassmann, it must have a property, not
typical of Grassmann coordinates, to commute with not only bosonic but 
also fermionic fields of the hierarchy, as follows from the continuum 
limit of relation \p{rule1}.

It is obvious that in addition to the coordinates $\pi$ and $p$, the phase
space must include a continuum ``lattice" coordinate $s$, Eq. \p{coord1}.
The resulting phase superspace $\{\pi, ~p,~ s\}$ of the dispersionless
$N{=}(1|1)$ 2DTL hierarchy contains the phase space $\{p,~ s\}$ of the 
dispersionless 2DTL hierarchy as a subspace. 

Having established the coordinates on the phase space, we can construct 
the Poisson superbracket for them. Recalling that the Poisson superbracket 
must agree with the superalgebra of the $Z_2$-graded operators    
$\{\sqrt{\hbar}e^{{\partial}}, ~e^{2{\partial}},~ {\hbar} j\}$, to which 
the phase superspace coordinates $\{\pi,~p,~ s\}$ correspond, we can find 
these superbrackets relatively easily. We here give the Poisson 
superbrackets between two arbitrary functions  
%$\Bbb F_{1,2}\equiv\Bbb F_{1,2}(\pi,p,s)$ 
%block letters
${\bF}_{1,2}\equiv {\bF}_{1,2}(\pi,p,s)$
on the phase space obtained as explained above,   
\begin{eqnarray}
\Bigl\{{\bF}_1,{\bF}_2\Bigr\}&=& 2p~\Bigl(\frac{\partial
{\bF}_1}{\partial p} ~\frac{\partial {\bF}_2}{\partial s}-
\frac{\partial {\bF}_1}{\partial s} ~\frac{\partial
{\bF}_2}{\partial p}+ \frac{\partial {\bF}_1}{\partial \pi}~
\frac{\partial {\bF}_2}{\partial \pi}\Bigr)\nonumber\\
&+&\pi ~\Bigl(\frac{\partial {\bF}_1}{\partial \pi}~
\frac{\partial {\bF}_2}{\partial s}- \frac{\partial
{\bF}_1}{\partial s}~ \frac{\partial {\bF}_2}{\partial \pi}\bigr).
\label{poissbrack}
\end{eqnarray}

We note that after the transition to the new basis
$\{{\widetilde s}, \widetilde p,\widetilde \pi\}$ 
in the phase space given by the formulas  
\begin{eqnarray}
{\widetilde s}&:=&\frac{s}{2}, \quad \widetilde p := \ln p, \quad
{\widetilde \pi} := \frac{\pi}{\sqrt{2p}}, \nonumber\\
s&:=&2{\widetilde s}, \quad p := e^{\widetilde p}, \quad \pi :=
\sqrt{2}{\widetilde \pi} e^{\frac{\widetilde p}{2}},
 \label{poissbrack2}
\end{eqnarray}
Poisson superbrackets \p{poissbrack} become 
\begin{eqnarray}
\Bigl\{{\bF}_1,{\bF}_2\Bigr\}&=& \frac{\partial {\bF}_1}{\partial
\widetilde p} ~\frac{\partial {\bF}_2}{\partial \widetilde s}-
\frac{\partial {\bF}_1}{\partial \widetilde s} ~\frac{\partial
{\bF}_2}{\partial \widetilde p}+ \frac{\partial {\bF}_1}{\partial
\widetilde \pi} ~\frac{\partial {\bF}_2}{\partial \widetilde \pi},
\label{poissbrack1}
\end{eqnarray}
which corresponds to the canonical orthosymplectic structure of the phase
superspace.  

We now proceed to the next stage of deriving the Lax representation of the
dispersionless $N{=}(1|1)$ 2DTL hierarchy. Heuristic formulas for the 
symbols ${\cal L}^{{\pm}}$ and $({\cal L}^{{\pm}})^{n}_{*}$ 
of the Lax operators $L^{\pm}$ \p{laxs1} and of the composite operators 
$(L^{\pm})^{n}_{*}$ (\ref{rule2}--\ref{laxs2}) are 
\begin{eqnarray}
L^{{\pm}}~ &\rightarrow & ~ \frac{1}{{\sqrt\hbar}}~{\cal
L}^{{\pm}},  \nonumber\\
{\cal L}^{+}=\sum^{\infty}_{k=0} (u_{2k+1}+u_{2k}\pi)p^{-k},
&\quad &{\cal L}^{-}=\sum^{\infty}_{k=0}
(v_{2k-1}+v_{2k}\pi)p^{k-1}
\label{displaxs1}
\end{eqnarray}
and
\begin{eqnarray}
&&(L^{{\pm}})^{2m}_{*}~ \rightarrow ~({\cal L}^{{\pm}})^{2m}_{*},
\quad (L^{{\pm}})^{2m+1}_{*}~ \rightarrow ~
\frac{1}{{\sqrt\hbar}}~({\cal L}^{{\pm}})^{2m+1}_{*}, \nonumber\\
&&\quad \quad \quad ({\cal L}^{+})^{2m}_{*}~~~:=~
\sum^{\infty}_{k=0} (u^{(2m)}_{2k}+u^{(2m)}_{2k-1}\pi)p^{m-k}, \nonumber\\
&&\quad \quad \quad ({\cal L}^{-})^{2m}_{*}~~~:=~
\sum^{\infty}_{k=0} (v^{(2m)}_{2k}+v^{(2m)}_{2k+1}\pi)p^{k-m}, \nonumber\\
&&\quad \quad \quad ({\cal L}^{+})^{2m+1}_{*}:=~
\sum^{\infty}_{k=0}
(u^{(2m+1)}_{2k+1}+u^{(2m+1)}_{2k}\pi)p^{m-k},\nonumber\\
&&\quad \quad \quad ({\cal L}^{-})^{2m+1}_{*}:=~
\sum^{\infty}_{k=0}
(v^{(2m+1)}_{2k-1}+v^{(2m+1)}_{2k}\pi)p^{k-m-1}, \label{diplaxs2}
\end{eqnarray}
respectively. By definition, all the fields  
$\{u^{(m)}_{k},~v^{(m)}_{k}\}$ with $k<0$  must be set equal to zero. In 
obtaining these expressions we have used substitutions \p{displimit4} and 
\p{displax1}. We note that the above symbols are not commutative in 
general,                        
\begin{eqnarray}
({\cal L}^{{\alpha}})^{k}_{*} ({\cal L}^{{\beta}})^{m}_{*}=
(-1)^{km}(({\cal L}^{{\beta}})^{m}_{*})^{*(k)} (({\cal
L}^{{\alpha}})^{k}_{*})^{*(m)}, \quad {\alpha},{\beta} =+,-,
\label{dispprop3}
\end{eqnarray}
which is related to the atypical property of the Grassmann coordinate 
$\pi$ noted above (see the remark after Eq. \p{displax1}).

In Lax representations (\ref{laxreprs1}) and \p{laxreprs2} for the 
$N{=}(1|1)$ 2DTL hierarchy and in defining relations (\ref{rule2}), 
we finally replace the Lax operators with their symbols
(\ref{displaxs1}--\ref{diplaxs2}) and the generalized
graded bracket (\ref{brack}) with Poisson superbracket \p{poissbrack} 
in accordance with                           
\begin{eqnarray}
\Bigl[~ . ~, ~. ~\Bigr\} ~\rightarrow  ~
{\hbar}~\Bigl\{~.~,~.~\Bigr\} \label{dispbrac}
\end{eqnarray}
and then perform substitution \p{displimit2} and take limit 
\p{displimit1}. This gives the expressions
\begin{eqnarray}
D^{\pm}_n {\cal L}^{{\alpha}}&=& \mp {\alpha} (-1)^n \Bigr\{
{((({\cal L}^{\pm})^{n}_{*})}_{-{\alpha} })^{*}, {\cal
L}^{{\alpha}}\Bigl\}, \quad n \in {\bN}, \quad {\alpha} =+,-,
\label{displaxreprs1}
\end{eqnarray}
\begin{eqnarray}
D^{\pm}_n ({\cal L}^{{\alpha}})^{m}_{*}&=& \mp {\alpha} (-1)^{nm}
\Bigr\{ {((({\cal L}^{\pm})^{n}_{*})}_{-{\alpha} })^{*(m)}, ({\cal
L}^{{\alpha}})^{m}_{*}\Bigl\},\quad n,m \in {\bN},
\label{displaxreprs2}
\end{eqnarray}
\begin{eqnarray}
({\cal L}^{{\alpha}})^{2m}_{*}:= \Bigl(~\frac{1}{2}~\Bigl\{({\cal
L}^{{\alpha}})^{*},{\cal L}^{{\alpha}}\Bigr\}~\Bigr)^{m},\quad
({\cal L}^{{\alpha}})^{2m+1}_{*}:={\cal L}^{{\alpha}}~({\cal
L}^{{\alpha}})^{2m}_{*} \label{disprule2}
\end{eqnarray}
for the dispersionless $N{=}(1|1)$ 2DTL hierarchy. Direct calculations
confirm that obtained relations (\ref{displaxreprs2}--\ref{disprule2}) 
correctly reproduce dispersionless flows (\ref{rec1}--\ref{eqvv6-}) and 
are therefore the sought Lax representation for the dispersionless 
$N{=}(1|1)$ 2DTL hierarchy. 

%zdes' okonchil vchera

\section{ Conclusions}

We have proposed new bracket operation (\ref{brack}) with the properties 
in Eqs. (\ref{prop1}--\ref{jacobi}) on the space of graded operators with 
an involution; this bracket generalizes the graded commutator in  
superalgebras. We then obtained a new form of Lax representation
(\ref{laxreprs1}--\ref{rule2}) and \p{laxreprs2} 
for the two-dimensional $N{=}(1|1)$ supersymmetric Toda lattice
hierarchy in terms of the generalized graded bracket. We next used this
representation to construct semiclassical limit 
(\ref{displimit1}--\ref{displimit4}) of this hierarchy\,---\,the 
dispersionless $N{=}(1|1)$ Toda hierarchy in 
Eqs. (\ref{equ1+}--\ref{eqvv6-})\,---\,and its Lax representation
(\ref{displaxs1}--\ref{diplaxs2}), (\ref{displaxreprs1}--\ref{disprule2}) 
on the graded phase superspace with Poisson bracket \p{poissbrack}. 
Finally, we established bosonic symmetries \p{disp-flowss+-bos1}
of dispersionless $N{=}(1|1)$ supersymmetric Toda equation \p{disptoda}.

{}~

{\sl One of us (V.~K.) first became acquainted with A.~A.~Logunov 
while still a second-year student in the Physics Department of Moscow 
State University (1956) and over the decades since then have always 
felt his firm support in matters both within and outside science. All our 
best wishes, dear Anatolii Alekseevich!}